\documentstyle[pre,multicol,aps,epsfig]{revtex}

\def\be{\begin{equation}}
\def\ee{\end{equation}}

\newcommand{\ub}{{\rm ub}}
\newcommand{\bd}{{\rm b}}

\newcommand{\fig}{Fig.\ }
\newcommand{\eq}{Eq.\ }
\newcommand{\dx}{\frac{\partial}{\partial x}}
\newcommand{\piad}{\pi_{\rm ad}}
\newcommand{\tpi}{\tilde\pi_{\rm ad}}

\def\siml{\mathrel{\mathchoice {\vcenter{\offinterlineskip\halign{\hfil
$\displaystyle##$\hfil\cr<\cr\sim\cr}}}
{\vcenter{\offinterlineskip\halign{\hfil$\textstyle##$\hfil\cr
<\cr\sim\cr}}}
{\vcenter{\offinterlineskip\halign{\hfil$\scriptstyle##$\hfil\cr
<\cr\sim\cr}}}
{\vcenter{\offinterlineskip\halign{\hfil$\scriptscriptstyle##$\hfil\cr
<\cr\sim\cr}}}}}

\begin{document}

\title{Traffic of molecular motors through tube-like compartments}
\author{Stefan Klumpp and Reinhard Lipowsky}
\address{Max--Planck--Institut f\"ur Kolloid- und
  Grenzfl\"achenforschung, 14424 Potsdam, Germany} \date{\today}

\maketitle

\begin{abstract}
  The traffic of molecular motors through open tube-like compartments
  is studied using lattice models.  These models exhibit
  boundary-induced phase transitions related to those of the
  asymmetric simple exclusion process (ASEP) in one dimension. The
  location of the transition lines depends on the boundary conditions
  at the two ends of the tubes.  Three types of boundary conditions
  are studied: (A) Periodic boundary conditions which correspond to a
  closed torus--like tube.  (B) Fixed motor densities at the two tube
  ends where radial equilibrium holds locally; and (C) Diffusive motor
  injection at one end and diffusive motor extraction at the other
  end. In addition to the phase diagrams, we also determine the
  profiles for the bound and unbound motor densities using mean field
  approximations and Monte Carlo simulations.  Our theoretical
  predictions are accessible to experiments.
\end{abstract}

\section{Introduction}

Molecular motors are proteins that transform the free energy released
from chemical reactions into mechanical work.  In this article we
consider a special class of motor proteins, namely cytoskeletal motors
which perform directed walks along cytoskeletal filaments as reviewed
in \cite{Howard2001,Woehlke_Schliwa2000}. In the cell, these motors
have different functions related, e.g., to vesicle transport and cell
division. The best studied examples are kinesins, which walk along
microtubules, and certain types of myosins, which walk along actin
filaments.  After a certain walking time, such a motor unbinds from
its filament because its binding energy is finite and can be overcome
by thermal activation. For kinesins, this typically happens after
100--150 steps or after a walking time of about 1.2--1.8 seconds, see,
e.g., \cite{Vale__Yanagida1996,Thorn__Vale2000}.  In many motility
assays, the filaments are immobilized on a substrate and are in
contact with an aqueous solution. In such a situation, the unbound
motors diffuse in the surrounding fluid until they eventually reattach
to the same or another filament.

Recently, we have introduced lattice models to study the motors'
random walks, which consist of many diffusional encounters between
motors and filaments in open and closed compartments
\cite{lipo173,lipo174}.  If many motors are placed in such a
compartment, hard core exclusion between the motors has to be taken
into account, since the motors are strongly attracted to the binding
sites of the filaments, so that the filaments get overcrowded.  These
models are new variants of driven lattice gas models and exclusion
processes, for which the active processes which drive the particles
are localized to the filaments.

Lattice models of driven diffusive systems have been studied
extensively in the last years, see e.g.\ 
\cite{Katz__Spohn1984,Kolomeisky__Straley1998}. The simplest model is
the asymmetric simple exclusion process (ASEP) in one dimension, where
particles hop on a one-dimensional lattice with a strong bias towards
one direction (in the simplest case there are no backward steps at
all) and the only interaction of the particles is hard core exclusion,
i.e., steps to occupied lattice sites are forbidden. When coupled to
open boundaries, this simple model already exhibits a complex phase
diagram, see e.g.  \cite{Kolomeisky__Straley1998}, which we will
review below in some detail.  The first model for the 1--dimensional
ASEP was introduced more than 30 years ago by MacDonald {\it et al.}
\cite{MacDonald__Pipkin1968,MacDonald_Gibbs1969} in the context of
protein synthesis by ribosomes on messenger RNA (mRNA). At that time
it was solved using a mean field approach and used to explain results
of radioactive labeling experiments
\cite{Dintzis1961,Naughton_Dintzis1962,Winslow_Ingram1966} which
showed that protein synthesis gets slower as the ribosome moves on the
mRNA template. The model of MacDonald {\it et al.} explained this by
the steric hindrance between successive ribosomes along the mRNA
track.  Two years later the same model was discussed by Spitzer as a
simple example for interacting particles in probability theory
\cite{Spitzer1970}.  Since then, the asymmetric simple exclusion
process (and variants) has been studied extensively as a generic model
for non-equilibrium phase transitions
\cite{Krug1991,Schuetz_Domany1993,Derrida__Pasquier1993} and
interacting stochastic systems \cite{Liggett1999} as well as in other
applications such as traffic flow \cite{Popkov__Schuetz2001}.  Many
properties of the 1--dimensional ASEP are known exactly.

As mentioned, the lattice models for random walks of molecular motors
differs from the driven lattice gas models in an important way: Walks
of molecular motors are only 'driven' as long as the motor is attached
to a cytoskeletal filament, hence 'driving' is localized to one or
several lines. It can therefore be viewed as an ASEP which has the
additional property, that particles (molecular motors) can unbind from
the track with a small probability, diffuse in the surroundings and
reattach to the same or another filament. In more mathematical terms,
the ASEP is coupled to a {\em symmetric} exclusion process via
adsorption and desorption of particles onto filaments.

Boundary conditions play an important role in driven systems. This
becomes apparent, e.g., if one compares a tube--like system with
periodic boundary conditions with one with closed boundaries. In the
system with closed boundaries, a traffic jam of motors arises at one
end of the system and the current of motors bound to the filament is
balanced by diffusive currents of unbound motors as first shown in
\cite{lipo173}. With periodic boundary conditions, motors arriving at
the right end of the system just restart their walk from the left end
and a net current through the systems is obtained.

\begin{figure}[htbp]
  \begin{center}
    \leavevmode
    \psfig{file=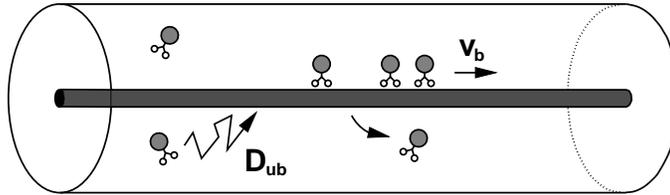,angle=0,width=.5\textwidth}
    \caption{Motor particles  which can bind and unbind to a filament (dark
     rod) within a cylindrical tube.  Single motors which are bound to the
     filament (and are not sterically  hindered by other motors)
     move with velocity $v_{\rm b}$ to the right. Unbound motors diffuse
     with diffusion coefficient $D_{\rm ub}$ in the surrounding liquid.}
    \label{fig:motor_tube_geom}
  \end{center}
\end{figure}

In this article, we study the stationary states of tube-like
compartments with open boundaries.  These compartments have the shape
of a cylinder and contain one filament which is placed along the
cylinder axis in order to obtain the simplest possible geometry. The
bound motors move along the filament and the unbound motors diffuse
within the cylinder, see \fig\ref{fig:motor_tube_geom}.  At the ends
of the tube, motors are inserted and extracted. Such a system is
accessible to {\it in vitro} experiments using standard motility
assays, but it can also be viewed as a strongly simplified model for
motor--based transport in an axon \cite{Goldstein_Yang2000}, if these
motors, which are synthesized in the cell body, are at least partly
degraded at the axon terminal \cite{Dahlstroem__Brady1991}, a
situation that can be mimicked by insertion and extraction of motors
at the ends of a tube.  The stationary states depend strongly on the
way, in which the motors are inserted and extracted at the boundaries
as we will explicitly demonstrate for three different types of
boundary conditions, see \fig\ref{fig:motor_tube_randbed}.

Our article is organized as follows. After introducing the model in
section \ref{sec:model}, we start in section \ref{sec:periodic_tube}
with periodic boundary conditions.  This case can be solved exactly,
since it satisfies local balance of currents in the radial direction,
see appendix \ref{app_RDB}. In section \ref{sec:open_tube}, we discuss
the situation in which the density of bound motors on the filament is
fixed at the boundaries. Finally, in section \ref{sec:boundDiff}, we
consider the case where the filament is shorter than the tube and the
motors diffuse into and out of the tube.  The main tool to study the
open systems are Monte Carlo simulations.  These are supplemented by
dynamical considerations and self-consistent or mean field
calculations.  Some details of the latter calculations are presented
in appendix \ref{app_mf}.

\section{Theoretical modelling}
\label{sec:model}

\subsection{Tube geometry}

We consider the motion of molecular motors in a cylindrical tube as
shown in \fig\ref{fig:motor_tube_geom}. The tube has length $L$ and
radius $R$.  The total number of motors within the tube, denoted by
$N_{\rm mo}$, defines the overall motor concentration
\begin{equation}
\rho_{\rm mo} \equiv \frac{N_{\rm mo}}{\pi R^2 L}
\quad .
\end{equation}
Note that this concentration corresponds to the particle number
density of the motors and, thus, is insensitive to the size of the
motor particles.  In general, the latter size depends on the type of
motor and on the type of cargo attached to it. In the following, we
will implicitly assume that the motor particle has a linear size which
is comparable to the basic length scale $\ell$ as defined further
below. If one wants to study the dependence on the motor particle size
in a systematic way, one should measure the overall motor
concentration in terms of the volume fraction of the motor particles.

The cylindrical tube contains one filament located along its symmetry
axis which is taken to be the x--axis; the two other Cartesian
coordinates are denoted by $y$ and $z$.  Motors bound to the filament
undergo directed motion, while unbound motors diffuse freely. Since
the motors are strongly attracted by the filament, a large fraction of
these motors is in the bound state and mutual exclusion from the
binding sites on the filament has to be taken into account even for
relatively small overall motor concentrations.

In order to include this mutual exclusion (or hard core repulsion) in
the theoretical description, we map the system onto a lattice gas
model on a simple cubic lattice. The lattice is oriented in such a way
that its three primitive vectors point parallel to the $x$--, $y$--,
and $z$--axis, respectively.  A rather natural choice for the lattice
parameter $\ell$ is the repeat distance of the filament, which is
$8\,$nm in the case of kinesin motors moving on microtubule and
$36\,$nm for myosin\,V motors moving on actin filaments.

The discretized tube consists of one line of binding sites, which
represents the filament, and $N_{\rm ch}$ unbound 'channels', i.e.\ 
lines of lattice sites parallel to the filament. Thus the cross
section $\phi$ of the tube is equal to \be \phi = (1+N_{\rm ch})\ell^2
\quad .  \ee For sufficiently large radii, one has $\phi\approx\pi
R^2$, while for small radii, there are corrections due to the
underlying lattice.

In the following, we will measure all distances such as $L$ and $R$ in
units of $\ell$.  Thus, for the simulations, both $L$ and $R$ will be
quoted as integers.  The integer value of $R$ corresponds to a certain
number of lattice sites along the Cartesian coordinates $y$ and $z$
which are perpendicular to $x$ and run parallel to two basis vectors
of the simple cubic lattice.  Thus, the interior of the tube contains
all channels with $y^2 + z^2 \le R^2$.

\subsection{Random walks with mutual exclusion}

At each time step, a bound motor attempts to make a forward or
backward step and to jump to the next lattice site to its right or to
its left with probability $\alpha$ or $\beta$. In addition, the motor
attempts to jump to each of the neighboring sites away from the
filament with probability $\epsilon/6$, and does not attempt to jump
at all, i.e., to rest at the filament site, with probability $\gamma$.
Since the sum of these probabilities is equal to one, we have
$\gamma=1-\alpha -\beta-2\epsilon/3 $. The velocity $v_\bd$ in the
bound state is given by $v_\bd=(\alpha-\beta)\,\ell/\tau$, where
$\tau$ is the basic time scale of these random walks.  In the
following, we will measure all times and rates in units of $\tau$ and
$\tau^{-1}$, respectively. Since backward steps are rare for
cytoskeletal motors, we will focus on the case $\beta=0$, i.e., we
will ignore backward steps.

If the particle unbinds from the filament, it attempts to jump to all
nearest neighbor sites of the simple cubic lattice with equal
probability $1/6$.  By choosing the time scale $\tau$ as
$\ell^2/D_\ub$, this hopping probability can be made to fit the
diffusion coefficient of unbound motors, $D_\ub$.  When measured in
units of $\ell$ and $\tau$, the dimensionless diffusion coefficient
$D_\ub = 1/6$. In principle, the resting probability $\gamma$ can then
be used to account for the ratio $D_\ub/(v_\bd\ell)$, which is quite
large in the case of kinesin, and to adapt the velocity $v_\bd$ to the
values obtained from experiments \cite{lipo173}.  For simplicity, we
will often choose $\gamma = 0$ in order to eliminate one parameter
from the problem.  If an unbound motor attempts to hop to a filament
site, the motor binds to it with sticking probability $\piad$, while
the step (and hence binding) is rejected with probability $1-\piad$.

Both in the bound and in the unbound state, hopping attempts can only
be successful, if the target site is not occupied by another motor;
otherwise the particle must stay where it is.  In the following, we
will mainly study overall motor concentrations $\rho_{\rm mo}$
in the range 
$0\leq \rho_{\rm mo} \protect\siml 0.05$. For such small values of
$\rho_{\rm mo}$, mutual exclusion of the unbound motors can be safely
ignored. However, because the motors are strongly attracted to the
filament, the concentration of bound motors is much larger and it is
crucial to take their mutual exclusion into account.

\section{Periodic boundary conditions}
\label{sec:periodic_tube}

First, we consider a cylindrical tube with periodic boundary
conditions in the longitudinal direction.  Because of the
translational invariance in the direction parallel to the filament,
there are no net radial currents in the stationary state. Indeed, a
non-zero radial current in a state, which is translationally invariant
in the longitudinal direction, would lead to net radial transport of
motor particles, which is incompatible with the reflecting radial
boundaries.  This means that there is a bound current $j_\bd$ on the
filament, but both currents of motors binding to and unbinding from
the filament {\em and radial currents of unbound motors } are balanced
locally. We call this situation {\em radial detailed balance} or {\em
  radial equilibrium}. If there is only one unbound channel, $N_{\rm
  ch}=1$, (or $N_{\rm ch}$ equivalent channels) radial equilibrium is
equivalent to adsorption equilibrium.  It is clear, that this will no
longer be true, if translational invariance is broken by boundaries or
blocked sites on the filament.  Another important property of systems
with periodic boundary conditions is that, in this case, the number
$N_{\rm mo}$ of motors in the system is conserved, which does not
apply to open systems.

Because of translational invariance along the x--axis, the bound and
unbound motor densities, $\rho_\bd$ and $\rho_\ub$, do not depend on
$x$. In addition, it follows from the absence of radial currents that
$\rho_\ub$ is also independent of the radial coordinate
$r=(y^2+z^2)^{1/2}$. Hence $\rho_\bd$ and $\rho_\ub$ are constant, and
radial equilibrium implies the relation
\begin{equation}
\label{RDB_Formel}
\frac{\epsilon}{6}\rho_\bd (1-\rho_\ub) =\frac{\piad}{6}\rho_\ub (1-\rho_\bd)
\end{equation}
which leads to
\begin{equation}\label{per_RB_bd}
\rho_\bd =\frac{\rho_\ub}{\epsilon/\piad+ (1-\epsilon/\piad)\rho_\ub}.
\end{equation}
Here and below, the densities $\rho_\bd$ and $\rho_\ub$ are local
particle number densities which satisfy \be 0 \le \rho_\bd \le 1
\qquad {\rm and} \qquad 0 \le \rho_\ub \le 1 \quad .  \ee In
dimensionful units, this corresponds to $0 \le \rho_\bd \le 1/\ell^3$
and $0 \le \rho_\ub \le 1/\ell^3$.

\begin{figure}[tbp]
\begin{center}
  \leavevmode
  \psfig{file=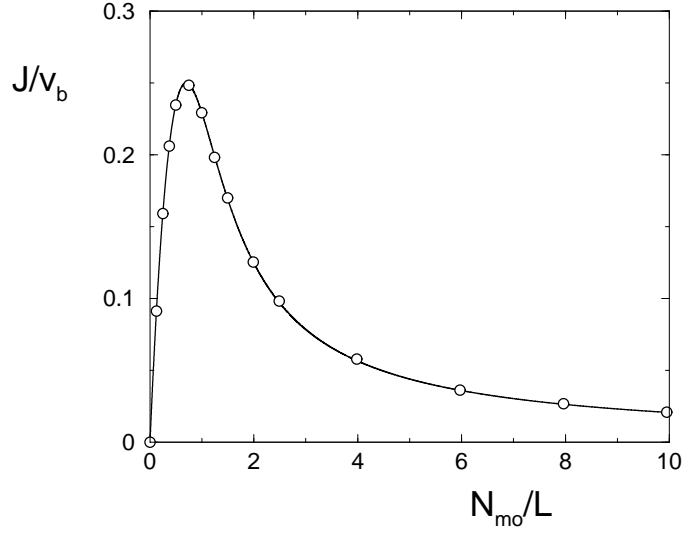,angle=-90,width=.5\textwidth}
  \parbox{\textwidth}{\caption{Reduced current $J/v_{\rm b}$ through the
      tube with periodic
      boundary conditions as a function of the reduced particle number
      $N_{\rm mo}/L$. The
      line is calculated from \eq(\ref{per_RB_analyt}), the Monte Carlo data
      are obtained for a tube of length $L=200$ and radius $R=25$
      corresponding to channel number $N_{\rm ch} = 1940$. The random walk
      probabilities are  $\beta = 0$,
      $\gamma = 99/100$, $\epsilon = 10^{-4}$, 
      $\alpha = 1 - \gamma - 2\epsilon/3$,
      and $\piad = 1$. }
\label{fig:strom_perRB}}
\end{center}
\end{figure}

Since the total number $N_{\rm mo}$ of motors is conserved, we can use
the normalization condition,
\begin{equation}
\rho_\bd +N_{\rm ch} \rho_\ub  =\frac{N_{\rm mo}}{L},
\end{equation}
to obtain a quadratic equation for the densities as a function of the
system size and the total number of motors. Since one root is always
negative, the physically meaningful solution is:
\begin{eqnarray}\label{per_RB_analyt}
\lefteqn{ \rho_\ub = \frac{1}{2N_{\rm ch} (1-\frac{\epsilon}{\piad})}
\Bigg[-1-N_{\rm ch}\frac{\epsilon}{\piad}+\frac{N_{\rm mo}}{L}
(1-\frac{\epsilon}
{\piad})  }\nonumber\\
 & & {} + \sqrt{\left(-1-N_{\rm ch}\frac{\epsilon}{\piad}+
\frac{N_{\rm mo}}{L}
(1-\frac{\epsilon}{\piad})\right)^2+4N_{\rm
ch}(1-\frac{\epsilon}{\piad})\frac{\epsilon}{\piad}
\frac{N_{\rm mo}}{L}}\, \Bigg].
\end{eqnarray}
From this expression for the unbound density, the bound density
$\rho_\bd$ follows via \eq(\ref{per_RB_bd}) and the stationary current
is given by $J=j_\bd=v_\bd\rho_\bd(1-\rho_\bd)$.  The current
calculated in this way is shown in \fig\ref{fig:strom_perRB} as a
function of $N_{\rm mo}/L$. The data points (circles) are the results
of Monte Carlo simulations for a system of length $L=200$ and radius
$R=25$ and are in very good agreement with the analytical solution.

It follows from the analytical solution that the current $J=j_\bd$
vanishes at $N_{\rm mo}/L=1+N_{\rm ch}$, behaves as
\begin{equation}
  \frac{J}{v_\bd}\approx \frac{1}{1+\frac{\epsilon}{\piad}N_{\rm ch}}
\frac{N_{\rm mo}}{L}
\end{equation}
for small $N_{\rm mo}/L$, and has the maximal value ${\rm
  max}(J/v_\bd)=1/4$ for $\rho_\bd=1/2$,
$\rho_\ub=\frac{\epsilon}{\piad}/(1+\frac{\epsilon}{\piad})$ and
$N_{\rm mo}/L=1/2+N_{\rm
  ch}\frac{\epsilon}{\piad}/(1+\frac{\epsilon}{\piad})$.

Let us add two remarks:

(i) \eq(\ref{RDB_Formel}), as stated here, can be considered as a mean
field equation. However, using the quantum Hamiltonian representation
of the stochastic process, it can be shown to hold exactly. The
calculation is simple, but rather technical and is therefore presented
in Appendix \ref{app_RDB}, which shows that stationary states can be
constructed as product measures provided the bound and unbound
densities satisfy \eq(\ref{RDB_Formel}).

(ii) Note that in contrast to a homogeneously driven lattice gas such
as the asymmetric simple exclusion process, there is no particle hole
symmetry here. Particles attempt to leave the filament with rate
$\epsilon/6$ to a neighboring site while holes do so with rate
$\piad/6$, i.e.\ particles are strongly attracted by the filament,
while holes are not. However, if one considers only the bound density,
the current density relationship $J=j_\bd=v_\bd\rho_\bd(1-\rho_\bd)$
is invariant under the exchange of particles and holes. As we will see
in the next section, this can lead to an apparent particle hole
symmetry for systems with radial equilibrium, because the radial
currents vanish and the state of the system can be determined by the
bound density alone.

\section{Open boundaries with radial equilibrium}
\label{sec:open_tube}

Now, let us consider the more interesting case, where the tube is open
and the densities at the left and right boundary are fixed. To be
precise, we consider two different sets of boundary conditions, (B)
and (C) as shown schematically in \fig\ref{fig:motor_tube_randbed}. In
this section, we study case (B) while case (C) will be considered in
the next section \ref{sec:boundDiff}.

For case (B), we add two layers of boundary sites at $x= 0$ and $x =
L+1$ with $y^2 + z^2 \le R^2$. As before, the filament is located at
$y = z = 0$. We then fix the density on the additional filament sites
according to \be \rho_\bd(x = 0) \equiv \rho_{\rm b,in} \qquad {\rm
  and} \qquad \rho_\bd(x = L+ 1) \equiv \rho_{\rm b,ex} \quad , \ee
see \fig\ref{fig:motor_tube_randbed}.  Furthermore, the densities on
the nonfilament boundary sites are chosen in such a way that radial
equilibrium as given by (\ref{RDB_Formel}) holds at both boundaries.

These boundary conditions are implemented by the following choice of
random walk probabilities.  First, we eliminate two parameters from
the problem, namely the jump probability $\beta$ to make backward
steps on the filament and the resting probability $\gamma$ to make no
step at all on the filament. Thus, we take $\beta = \gamma = 0$
throughout this section.

Next, when we choose a site within the left or right boundary layer
during the Monte Carlo sweep, we first draw a random number $\omega$
which is uniformly distributed over the interval $0 < \omega \le 1$.
The chosen left or right boundary site is taken to be occupied if
$\omega \le \rho_{\rm b,in}$ or $\omega \le \rho_{\rm b,ex}$,
respectively.  If a boundary site with $y^2 + z^2 > 0$ is occupied by
a motor particle, this particle attempts to jump into the tube with
probability $1/6$. If the left boundary site with $y = z = 0$ is
occupied, the corresponding particle attempts to jump onto the left
end of the filament with probability $(1 - 2 \epsilon/3)$. If the
right boundary site with $y = z = 0$ is occupied, this particle cannot
enter since $\beta = 0$.  In addition, all particles which jump from a
site within the tube, i.e., from a lattice site with $ 1 \le x \le L$,
onto a boundary site, are extracted from the tube.

\subsection{Phase diagram}
\label{sec:boundRDB}

In the limiting case with $N_{\rm ch}=0$ and $\epsilon = 0$, our
system becomes equivalent to the 1--dimensional ASEP, for which the
density profiles and the phase diagram are known exactly
\cite{Schuetz_Domany1993,Derrida__Pasquier1993}. Let us therefore
summarize some of the known properties of this process, for which we
use symbols without the subscript '$\bd$'.

For the 1--dimensional ASEP, there are three different phases.
If the density $\rho_{\rm in}$ at the left boundary is small and
satisfies $\rho_{\rm in}<1/2$ and if the density $\rho_{\rm ex}$ at
the right boundary is not too large with $\rho_{\rm ex}<1-\rho_{\rm
  in}$, the system is in the low density (LD) phase, for which the
bulk density $\rho^0$ is equal to the left boundary density, and the
current is given by $J=v \rho_{\rm in}(1-\rho_{\rm in})$.  Because of
the particle hole symmetry, an analogous situation holds for
$\rho_{\rm ex}>1/2$ and $\rho_{\rm in}>1-\rho_{\rm ex}$ [high density
(HD) phase]. Now, the bulk density is given by $\rho^0=\rho_{\rm ex}$
and the current is $J=v \rho_{\rm ex}(1-\rho_{\rm ex})$. At the line
$\rho_{\rm ex}=1-\rho_{\rm in}<1/2$, a discontinuous phase transition
takes place, and the bulk density jumps from the left boundary value
to the right boundary value.  Finally, for $\rho_{\rm in}>1/2$ and
$\rho_{\rm ex}<1/2$, the bulk density $\rho^0=1/2$ and the current
attains its maximal value $J=v/4$. Therefore, this phase is called
maximal current (MC) phase.  The phase transition towards the maximal
current phase is continuous with a diverging correlation length.

The formation of three different phases can be understood in terms of
the underlying dynamics of domain walls and density fluctuations
\cite{Kolomeisky__Straley1998}. In the low density and high density
phases, the selection of the stationary state is governed by domain
wall motion. Thus consider a domain wall, which forms between regions
of different densities. If its velocity is positive, the domain wall
travels to the right boundary and the bulk density is equal to the
left boundary density (low density phase). Likewise, the bulk density
is given by the right boundary density, if the domain wall velocity is
negative (high density phase). At the transition line with $\rho_{\rm
  ex}=1-\rho_{\rm in}<1/2$ the domain wall velocity is zero, and
domain walls diffuse through the system.

The second mechanism is related to density fluctuations. If a small
perturbation of the density, corresponding, e.g., to some added
particles, enters the system at the left boundary, it can move with
positive or negative velocity, i.e., it can spread into the bulk or is
driven back towards the boundary. In the maximal current phase the
velocity of such density perturbations, is negative and density
fluctuations coming from the left boundary are driven back to the
boundary.  Hence, increasing $\rho_{\rm in}$ does not increase the
bulk density, since additional particles cannot enter the system
(overfeeding effect). Both the velocities of domain walls and of
density fluctuations are governed by the same density current relation
which is $j=v\rho(1-\rho)$ for the 1--dimensional ASEP
\cite{Kolomeisky__Straley1998}.

For the filament in a tube as considered here, the velocities of
domain walls and density fluctuations are similar to those for the
ASEP in one dimension.  This can be understood from the fact that the
density--current relationship on the filament is the same as for the
1--dimensional ASEP provided one rescales all currents by the factor
$(1 - 2 \epsilon/3)$ which arises from the possibility to unbind from
the filament.

Thus, let us first consider the case, where the behavior on the
filament is determined by domain walls, i.e.\ the low density and high
density phases.  Drift motion of domain walls is governed by the
domain wall velocity on the filament, which is $v_s=v (1-\rho_{\rm
  ex}-\rho_{\rm in})$ for the one-dimensional ASEP
\cite{Kolomeisky__Straley1998}. In the tube system considere here, the
domain wall velocity is slowed down compared to this value, because
the domain wall of the unbound density must follow the bound density
domain wall by binding and unbinding of motors to and from the
filament. However, the sign of the domain wall velocity is the same
for this tube system as for the one--dimensional ASEP, and the domain
wall velocity changes sign at the same values of the boundary
densities. An explicit expression for $v_s$ can be obtained from the
general expression for the domain wall velocity as given in Ref.\ 
\cite{Kolomeisky__Straley1998} by integrating the density over the
tube cross-section, which leads to
\begin{equation}
v_s=\frac{v_\bd\rho_{\rm b,ex}(1-\rho_{\rm b,ex})-v_\bd\rho_{\rm b,in}
(1-\rho_{\rm b,in})}{\rho_{\rm b,ex}-\rho_{\rm b,in}+N_{\rm ch}\rho_{\rm
ub,ex}-N_{\rm ch}\rho_{\rm ub,in}}
\end{equation}
for the geometry considered here.  Remember that the unbound densities
are related to the bound densities by radial equilibrium at the
boundaries.  If a domain wall spreads from the left or right boundary
into the system, radial equilibrium will hold approximately in the
bulk because of translational invariance, but, in addition, all the
way down to the dominating boundary, where we have imposed radial
equilibrium via the boundary conditions.  Therefore, the current and
the bulk density are determined by the bound density at the
boundaries.

The drift velocity of fluctuations of the bound density is
$v_c=1-2\rho_\bd$ \cite{Kolomeisky__Straley1998}.  If the behavior on
the filament is determined by this velocity, i.e.\ in the case
$\rho_{\rm b,in}>1/2$ and $\rho_{\rm b,ex}<1/2$, radial equilibrium
will not hold up to the boundary and unbound motors can enter the
tube.  However, after a short distance they will bind to the filament
and act like an additional particle in the maximal current phase: The
density fluctuation generated by the additional particle moves towards
the boundary. Therefore in the bulk, the bound density is 1/2 and the
current is $v_\bd/4$.

\begin{figure}[tbp]
\begin{center}
  \leavevmode
  \psfig{file=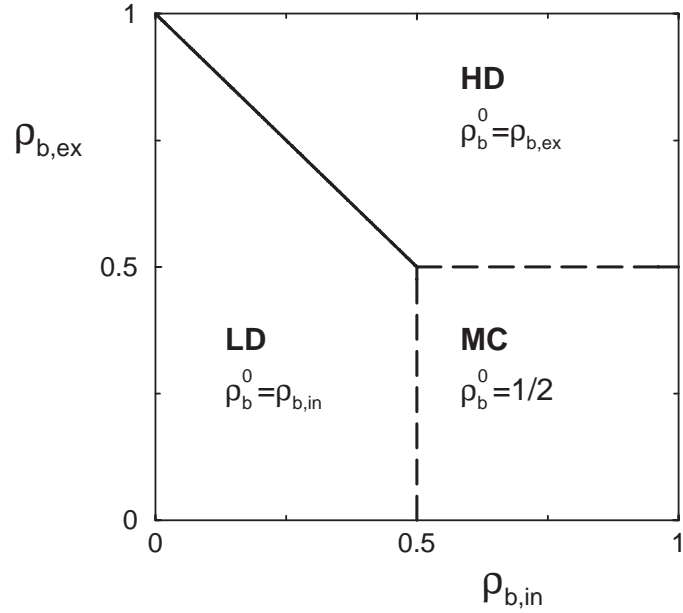,angle=-90,width=.5\textwidth}
  \parbox{\textwidth}{\caption{Phase diagram for
      motor traffic in open tubes with  boundary condition (B) as a function of
      the boundary densities $\rho_{\rm b,in}$ and $\rho_{\rm b,ex}$
      at the left and   right end of the filament, respectively. There are
      three
      phases distinguished by the bulk value of the bound density
      $\rho_\bd^0$:
      a low density phase (LD), a high density phase (HD), and a maximal
      current
      phase (MC). This phase diagram is identical to the phase diagram of
      the 1--dimensional ASEP as explained in the text.}
\label{fig:phasendiag_ASEP}}
\end{center}
\end{figure}

Summarizing these considerations, we predict to find exactly the same
phase diagram as for the 1--dimensional ASEP, see
\fig\ref{fig:phasendiag_ASEP}.  In fact, we have chosen our boundary
conditions (B) in such a way that we only have to replace the density
$\rho$ of the 1--dimensional ASEP by the bound density $\rho_\bd$ and
the boundary densities $\rho_{\rm in}$ and $\rho_{\rm ex}$ by the
boundary densities on the filament, $\rho_{\rm b,in}$ and $\rho_{\rm
  b,ex}$.  The unbound density in the bulk is obtained by radial
equilibrium from the bound density. These expectations are confirmed
(i) by a detailed analysis of the discrete mean field equations and
(ii) by extended Monte Carlo simulations.

A remarkable feature of the phase diagram is that it seems to exhibit
particle hole symmetry, while the dynamics does not. The signature of
particle hole symmetry in the phase diagram is the symmetry between
the high density and low density phase.  In both these phases, the
bulk density is approximately constant and radial equilibrium holds
approximately in the whole system except close to the left or right
boundary. Radial equilibrium holds exactly at that boundary which
determines the bulk behavior. Therefore radial currents vanish on
average and the phase diagram is determined by the bound density
alone, resulting in a phase diagram which gives the impression of
particle hole symmetry, although this symmetry is broken. Indeed, even
though the substitution of the bound density $\rho_\bd$ by
$1-\rho_\bd$ leads to the same bound current $j_\bd$, it does not lead
to the unbound density $1-\rho_\ub$. Two stationary states with bound
densities that are related through particle hole symmetry are
characterized by two unbound densities which are both smaller than the
bound ones and thus break the apparent symmetry.

\subsection{Density profiles}
\label{sec:profile}

To discuss the concentration profiles of the bound and unbound motors,
we use continuum mean field equations and compare the mean field
results with simulations. In addition, we make a two-state
approximation, i.e., we consider the case of a single unbound channel,
so that a motor can be in only two states, bound to the filament or
unbound. The two-state model is exact for an arbitrary number $N_{\rm
  ch}$ of \emph{equivalent} unbound channels, but it can also serve as
an approximation for the original tube systems: Since the unbound
density depends only weakly on the radial coordinate, we will consider
the approximation in which $\rho_\ub$ is taken to be independent of
$r$ and the unbound channels are taken to be equivalent.  This
approximation corresponds to a two--state model in which the bound
motors are described by the density $\rho_\bd(x)$ and the unbound
motors by the density $\rho_\ub(x)$. The effective diffusion
coefficient for the unbound motors is given by $N_{\rm ch}D_\ub$.

Using again the mean field approximation, the total current $J$
parallel to the tube axis is now given by
\begin{equation}
  \label{mf1}
  J= v_\bd \rho_\bd(1-\rho_\bd) - D_\bd \dx\rho_\bd - N_{\rm ch} D_\ub
  \dx\rho_\ub
  \label{E.TwoState1}
\end{equation}
and the equality between incoming and outgoing currents at any lattice
site leads to
\begin{equation}
  \label{mf2}
  \dx \Big[ v_\bd \rho_\bd(1-\rho_\bd) - D_\bd \dx\rho_\bd \Big]=
   \tpi\rho_\ub(1-\rho_\bd)-\tilde\epsilon\rho_\bd(1-\rho_\ub),
   \label{E.TwoState2}
\end{equation}
where $D_\bd$ and $D_\ub$ are the diffusion coefficients of motors in
the bound and unbound state, respectively. In addition, we have
introduced the rescaled binding and unbinding rates,
$\tpi=\frac{2}{3}\piad$ and $\tilde\epsilon=\frac{2}{3}\epsilon$.

In order to determine the density profiles far from the boundaries, we
first calculate the homogeneous, $x$--independent solutions,
$\rho_\bd^0$ and $\rho_\ub^0$, of the two mean field equations
(\ref{E.TwoState1}) and (\ref{E.TwoState2}). The first equation
(\ref{E.TwoState1}) for the total current $J$ then reduces to the
current--density relationship
\begin{equation}
J = v_\bd \rho_\bd^0(1-\rho_\bd^0)
\end{equation}
whereas the second equation (\ref{E.TwoState2}) becomes
\begin{equation}
\tpi\rho_\ub^0(1-\rho_\bd^0) = \tilde\epsilon\rho_\bd^0(1-\rho_\ub^0)
\end{equation}
which implies radial equilibrium for the $x$--independent solutions.
We then decompose the densities according to
\begin{equation}
\rho_\bd(x)=\rho_\bd^0+\eta_\bd(x) \qquad {\rm and} \qquad
\rho_\ub(x)=\rho_\ub^0+\eta_\ub(x),
\label{E.DensityDecomposition}
\end{equation}
and expand the mean field equations (\ref{E.TwoState1}) and
(\ref{E.TwoState2}) in powers of the density deviations $\eta_\bd$ and
$\eta_\ub$. The details of this expansion are described in Appendix
\ref{app_mf}.

\subsubsection{Low density and high density phases}

For the high and low density phases with $\rho_\bd^0\neq 1/2$, the
expansion of the mean field equations up to first order in $\eta_\bd$
and $\eta_\ub$ leads to an exponential approach $\sim \exp(x/\xi)$ of
the density profiles towards the homogeneous solutions $\rho_\bd^0$
and $\rho_\ub^0$, see Appendix \ref{app_mf}. The corresponding decay
length $\xi$ satisfies the cubic equation
\begin{equation}
\label{E.xi}
 - v_\bd(1-2\rho_\bd^0) \xi^3 +
  (D_\bd+ N_{\rm ch} D_\ub g) \xi^2 +N_{\rm ch}
D_\ub\frac{v_\bd(1-2\rho_\bd^0)}{A} \xi
  -N_{\rm
ch} D_\ub\frac{D_\bd}{A}  = 0
\end{equation}
which is solved numerically.

For the ASEP in one dimension, one has $N_{\rm ch} = 0$ and
(\ref{E.xi}) reduces to $- v_\bd(1-2\rho_\bd^0) \xi^3 + D_\bd \xi^2 =
- \xi^2(\xi - \xi_0) = 0$ with
\begin{equation}
\label{E.xi0}
\xi_0 \equiv D_\bd / v_\bd(1-2\rho_\bd^0)
\quad .
\end{equation}
Thus, in this limit, mean field theory leads to the correlation length
$\xi = \xi_0$. For $N_{\rm ch} > 0$, we choose the unique solution of
(\ref{E.xi}) which approaches (\ref{E.xi0}) as $N_{\rm ch}$ vanishes.
This solution behaves as $ \xi \approx \xi_0 (1 + g N_{\rm ch} D_\ub
/D_\bd)$ for small $N_{\rm ch}$.

\begin{figure}[tbp]
\begin{center}
  \leavevmode
  \psfig{file=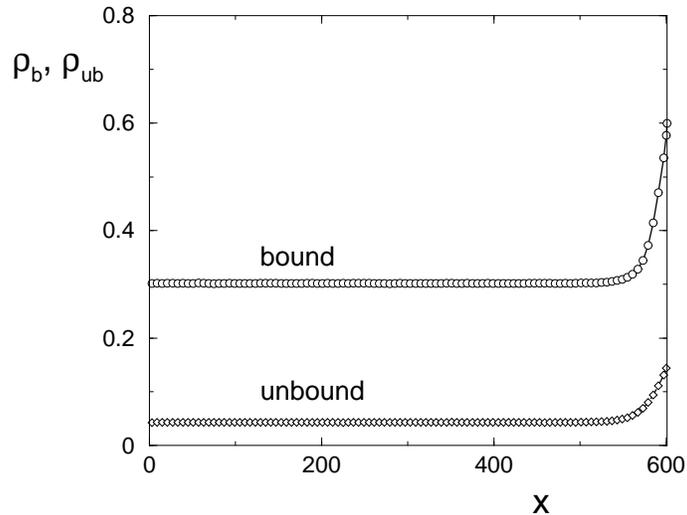,angle=-90,width=.5\textwidth}
  \parbox{\textwidth}{\caption{Bound  and unbound
      density profiles
      $\rho_\bd$ and $\rho_\ub$ as a function of the spatial coordinate $x$
      for the low density phase with boundary conditions (B) and boundary
      densities  $\rho_{\rm b,in}=0.3$ and $\rho_{\rm b,ex}=0.6$.
      The unbound density has been averaged over the
      tube cross section and multiplied by a scale factor of 10. All data
      points
      are averages over six lattice sites in direction parallel to the
      filament. The tube has length $L = 600$ and radius $R=10$. The random
      walk probabilities are  $\beta=\gamma=0$,
      $\epsilon = 10^{-2}$, $\alpha = 1 - 2\epsilon/3$, and $\piad=1$. }
\label{fig:profile_LD}}
\end{center}
\end{figure}

\begin{figure}[tbp]
\begin{center}
  \leavevmode
  \psfig{file=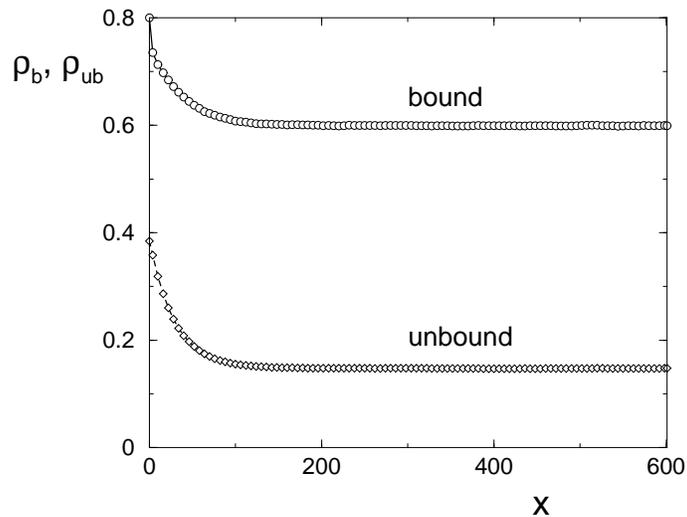,angle=-90,width=.5\textwidth}
  \parbox{\textwidth}{\caption{Bound   and unbound
        density profiles  $\rho_\bd$ and $\rho_\ub$
      as functions of the spatial coordinate $x$ for the high density phase;
       boundary
      conditions (B) with $\rho_{\rm b,in}=0.8$ and $\rho_{\rm b,ex}=0.6$.
      The unbound density
      has been averaged over the
      tube cross section and multiplied  by a scale factor of 10. The geometry
      and the random walk probabilities are the same as in
      \fig\ref{fig:profile_LD}.}
\label{fig:profile_HD}}
\end{center}
\end{figure}

In addition to the mean field calculation, we again used Monte Carlo
simulations in order to determine the density profiles as shown in
Figs.\,\ref{fig:profile_LD} and \ref{fig:profile_HD}.  As predicted by
the mean field calculation, the constant bulk densities for the bound
and unbound states are approached exponentially in the low and high
density phases. The corresponding decay length $\xi$ is found to be
the same for the bound and the unbound density and to diverge as one
approaches the maximal current phase.

\begin{figure}[htbp]
  \begin{center}
    \leavevmode
    \psfig{file=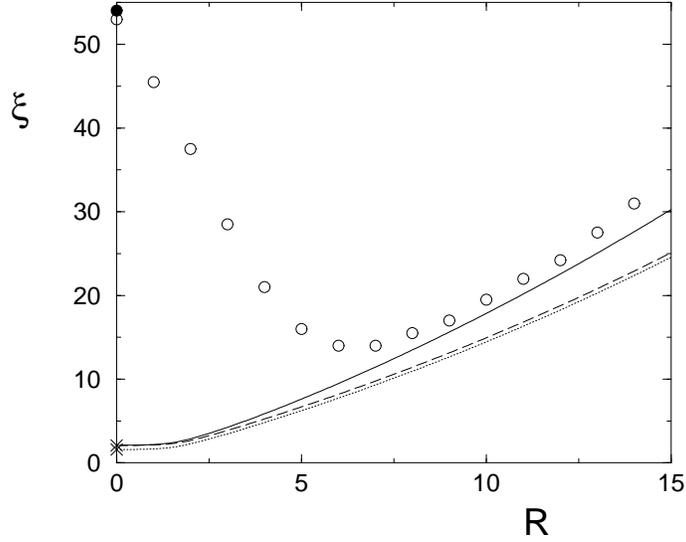,angle=-90,width=.5\textwidth}
    \caption{Localization length $\xi$ as a function of the tube
      radius $R$ for boundary conditions (B) with $\rho_{\bd,\rm in}=0.38$ and
      $\rho_{\bd,\rm ex}=0.6$.
      Circles are simulation data, obtained for a tube with length $L=600$
      and with random walk probabilities as in \fig\ref{fig:profile_LD};
      lines are the corresponding results of mean field
      calculations (dotted: discrete two-state approximation, dashed:
      continuous two-state approximation, solid: full diffusion
      equation, see text). The two crosses and the filled circle at $R=0$
      represent
      the mean field results and the exact result  for
      the 1--dimensional ASEP, respectively.}
    \label{fig:kurve_l_R}
  \end{center}
\end{figure}

Some simulation results for the decay length $\xi$ are displayed in
\fig\ref{fig:kurve_l_R}. In this case, the radius $R$ of the tube was
varied while the boundary densities were kept fixed.  The latter
densities were chosen to be $\rho_{\rm b,in} = 0.38$ and $\rho_{\rm
  b,ex} = 0.6$ which lies within the low density phase but is close to
the phase transition line which separates the low from the high
density phase. One surprising feature of the Monte Carlo data for the
decay length $\xi$ is that they exhibit a pronounced {\em minimum} as
a function of tube radius $R$.

For comparison, we also display in \fig\ref{fig:kurve_l_R} the
$\xi$--values as obtained from several mean field approximations
corresponding to the dashed, dotted and solid lines.  The dashed line
is obtained from the solution of equation (\ref{E.xi}) which we
derived from the continuous mean field approximation for the
two--state model. We also determined this quantity using a lattice
version of this approximation (dotted line) and a more elaborate mean
field approximation (solid line) in which we solved the diffusion
equation in the cylindrical compartment and matched this solution to
the directed transport along the filament.

Inspection of \fig\ref{fig:kurve_l_R} shows that all three mean field
approximations are quite consistent with each other and lead to $\xi
\sim R^2$ as follows from (\ref{E.xi}). Such an increase of $\xi$ for
large $R$ is in fair agreement with the Monte Carlo data displayed in
\fig\ref{fig:kurve_l_R}.  However, in contrast to the Monte Carlo
simulations, all three mean field approximations give a {\em
  monotonic} increase of $\xi$ with increasing $R$.

The largest discrepancy between the mean field results and the Monte
Carlo data is found in the limit of small $R$ for which one recovers
the 1--dimensional ASEP.  In this limit, the Monte Carlo data should
be quite reliable as one concludes from the value obtained for $R = 0$
which is in very good agreement with the exact solution for the
1--dimensional ASEP as given by \cite{Schuetz_Domany1993}
\begin{equation}
\xi = \left| \frac{1}{\xi_{\rm in}} + \frac{1}{\xi_{\rm ex}} \right|^{-1}
\quad {\rm with} \quad \xi_k \equiv - \frac{1}{\ln [4 \rho_k (1 - \rho_k)]}
\end{equation}
for $k = $ in, ex. Thus, we conclude that the decay length $\xi$ does
indeed exhibit a nonmonotonic dependence on the tube radius $R$ and
that this behavior is not reproduced by the mean field approximation.

Finally, we note that the different behavior of the decay length $\xi$
for large and for small $R$ is correlated with a qualitatively
different behavior of the corresponding density profiles as observed
in the Monte Carlo simulations.  As an example, let us consider the
density profiles within the low density phase as in
\fig\ref{fig:profile_LD}. In this case, the bound and unbound
densities exhibit plateau regions which are determined by their values
$\rho_{\rm b,in}$ and $\rho_{\rm ub,in}$ at the left boundary.  As one
gets closer to the right boundary where the motors can leave the tube,
the densities start to deviate from these constant values, and these
deviations grow exponentially as $\sim \exp(x/\xi)$.  For small $R$,
the corresponding profiles are convex upwards for all values of $x$.
For large $R$, on the other hand, the profile exhibits an inflection
point close to the right boundary.  This inflection point moves
towards the interior of the tube as $R$ is further increased.

\subsubsection{Maximal current phase}
\label{sec:profileMC}

In the maximal current phase, one has $\rho_\bd^0=1/2$, and one has to
consider terms up to second order in the density deviations $\eta_\bd$
and $\eta_\ub$, see Appendix \ref{app_mf}. One then finds that
$\eta_\bd$ satisfies the nonlinear differential equation
\begin{equation}
\label{mf_rhobd_tube}
  (D_\bd+ g N_{\rm ch} D_\ub)\dx\eta_\bd \approx -v_\bd\eta_\bd^2
\end{equation}
with
\begin{equation}\label{def_Gamma}
  g \equiv \frac{\tilde\epsilon
(1-\rho_\ub^0)+\tpi\rho_\ub^0}{\tpi(1-\rho_\bd^0)+\tilde\epsilon\rho_\bd^0}.
\end{equation}
This differential equation can be solved by separation of variables.
The solution behaves as
\begin{equation}
  \eta_\bd \approx \frac{D_\bd+ g \,N_{\rm ch} D_\ub}{v_\bd\,x}
\qquad {\rm for} \quad {\rm large} \quad x
\end{equation}
i.e., the deviation of the bound density from its asymptotic bulk
value $\rho_\bd^0=1/2$ decays as $\sim 1/x$ within the mean field
approximation.

For $N_{\rm ch} = 0$, this becomes identical with the mean field
solution for the ASEP in one dimension as discussed in
\cite{Krug1991}. In this latter case, an exact solution is available
\cite{Schuetz_Domany1993,Derrida__Pasquier1993}, which shows that the
density profile decays as $\sim 1/x^{1/2}$, i.e., with a different
exponent.  This is related to the fact that density fluctuations
spread superdiffusively in the 1--dimensional ASEP
\cite{vanBeijeren__Spohn1985}.  The dispersion $\langle (\Delta x)^2
\rangle$ of an ensemble of particles in such a system behaves as $
\langle (\Delta x)^2 \rangle \sim t^{4/3}$ for large times $t$.  This
superdiffusive spreading of density fluctuations in one dimension can
be taken into account within a mean field approximation if one
considers a scale-dependent diffusion coefficient as shown in Ref.
\cite{Krug1991}.  In the following, we will extend this approach to
the tube geometry considered here.

Thus, we now replace $D_\bd$ in the mean field equation
(\ref{mf_rhobd_tube}) by a scale--dependent diffusion coefficient $
D_\bd(x)$ and consider the modified mean field equation
\begin{equation}
 [D_\bd(x) + g N_{\rm ch} D_\ub]\dx\eta_\bd \approx -v_\bd\eta_\bd^2
\quad .
\label{E.meanfieldmodified}
\end{equation}
A convenient choice for $ D_\bd(x)$ which embodies the correct
superdiffusive behavior for the ASEP in one dimension is
\begin{equation}
\label{scale_dep_D_b}
  D_\bd(x) \equiv D_{sc} \left( 1+\sqrt{ \frac{x - x_0}{ x_{sc}}  } \right)
\quad .
\end{equation}
The left boundary is located at $x = x_0$, and $D_{sc}$ and $x_{sc}$
represent two scale parameters.  With this choice, the modified mean
field equation (\ref{E.meanfieldmodified}) can again be solved by
separation of variables. As a result, we obtain
\begin{equation}
\label{E.bdensitydev}
 \eta_\bd(x) = a \left[ \frac{a}{\eta_\bd(x_0)} +
\sqrt{y}  - b \ln( 1 +  \sqrt{y}/b) \right]^{-1}
\end{equation}
where we have introduced the abbreviations
\begin{equation}
\label{E.yDefined}
y \equiv (x - x_0) / x_{sc}
\quad ,
\end{equation}
\begin{equation}
\label{E.aDefined}
a \equiv D_{sc} / 2 v_b x_{sc}
\quad ,
\end{equation}
and
\begin{equation}
b \equiv 1 + g \,N_{\rm ch} D_\ub/D_{sc}
\quad .
\end{equation}
The 'initial' value $\eta_\bd(x_0) = \rho_{\rm b,in} - \frac{1}{2}$
denotes the density deviation at the left boundary.

Far from the left boundary, i.e., for large values of $y \sim x$, the
expression (\ref{E.bdensitydev}) leads to the asymptotic behavior
\begin{equation}
 \eta_\bd(x) \approx \frac{a}{\sqrt{y}}
\left[ 1 + b \frac{\ln(\sqrt{y})}{\sqrt{y}} \right]
\quad .
\label{E.bdensitydev1}
\end{equation}
Thus, the deviation of the bound density from its asymptotic value now
decays as $1/\sqrt{y} \sim 1/\sqrt{x}$ as for the ASEP in one
dimension.

In addition, we also obtain a correction term in
(\ref{E.bdensitydev1}) which depends on $b = 1 + g \,N_{\rm ch}
D_\ub/D_{sc}$.  For large tube radius $R$, one has $N_{\rm ch} \sim
R^2$ and $b \sim N_{\rm ch} \sim R^2$. Therefore, the correction term
becomes large for large $R$. We can now define a crossover length $y =
y_*$ at which the correction term has the same size as the leading
term. This leads to the implicit equation
\begin{equation}
\sqrt{y_*} / \ln(\sqrt{y_*}) =  b
\end{equation}
and, thus, to
\begin{equation}
y_* \approx [b \ln(b) ]^2 \sim [R^2 \ln(R^2)  ]^2
\qquad {\rm for} \quad {\rm large} \quad b \sim R^2
\quad .
\label{E.CrossoverLength}
\end{equation}

Close to the left boundary, i.e., for small values of $y \sim x -
x_0$, the expression (\ref{E.bdensitydev}) for the deviation of the
bound density from its asymptotic value $\rho_\bd^0=1/2$ leads to
\begin{equation}
 \eta_\bd(x)  \approx \eta_\bd(x_0) [ 1 - \frac{1}{2 a b} \eta_\bd(x_0)
y ].
\label{E.bdensitydev2}
\end{equation}
We may now define a second crossover length or extrapolation length $y
= y_{**}$ at which the two terms in (\ref{E.bdensitydev2}) cancel.
This extrapolation length is given by
\begin{equation}
y_{**} = 2 a b / \eta_\bd(x_0) = 2 a b / (\rho_{\rm b,in} - \frac{1}{2})
\quad .
\label{E.ExtrapolationLength}
\end{equation}
For large tube radius $R$, the latter length scale grows as $y_{**}
\sim b \sim R^2$. In general, the extrapolation length can have both
signs but, for the maximal current phase, one always has $\rho_{\rm
  b,in} > 1/2$ and, thus, $y_{**} > 0$.

There is also an intermediate range of $y$--values defined by $y_{**}
\ll y \ll b^2 \sim y_*$. For these $y$--values, the deviation
$\eta_\bd$ of the bound density from its asymptotic value as given by
(\ref{E.bdensitydev}) simplifies and becomes \be \eta_\bd \approx
\frac{2 a b }{y} = \frac{D_{sc} + g \,N_{\rm ch} D_\ub}{v_\bd x_{sc}}
\, \frac{1}{y} \quad .  \ee Thus, for these intermediate $y$--values
the density deviation decays again as $\sim 1/y$.

In summary, the theory described here indicates (i) that the 'initial'
value at the left boundary is felt up to an extrapolation length
$y_{**} \sim b \sim R^2$, (ii) that the true asymptotic behavior of
the density deviation is obtained for $y > y_* \sim R^4$ as follows
from (\ref{E.CrossoverLength}), and (iii) that the density deviation
decays as $1/y$ on intermediate length scales with $y_{**} \ll y \ll
y_*$.

\begin{figure}[htbp]
  \begin{center}
    \leavevmode
    \psfig{file=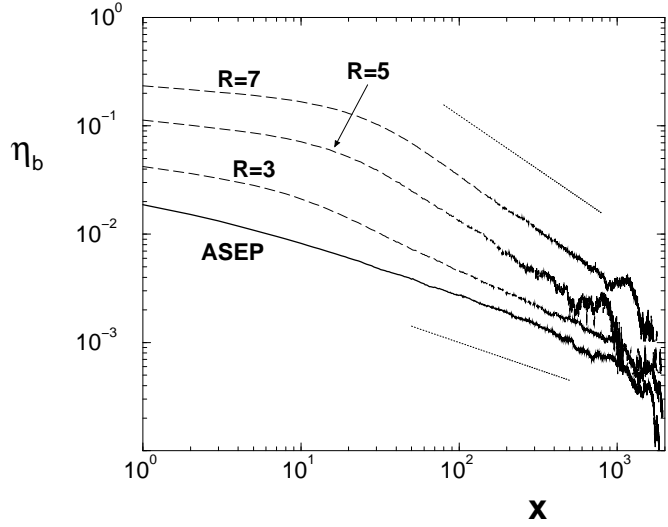,angle=-90,width=.5\textwidth}
    \parbox{\textwidth}{\caption{Deviation $\eta_\bd$ of the bound density
        from its constant value far from the boundaries as a function
        of the spatial coordinate $x$ for the 1--dimensional ASEP (solid
        line) and for motor traffic in tubes with radius $R=3,5,7$ (dashed
        lines) and
         boundary conditions (B) with $\rho_{\rm b,in}=0.8$ and
        $\rho_{\rm b,ex}=0.5$.   The curves for the 1--dimensional ASEP and
        for the tube radii
        $R=3,5$ have been multiplied by scale factors   0.1, 0.2, and 0.5,
        respectively.
        The thin dotted lines correspond to the decay laws $\sim 1/x $ and
        $\sim 1/ \sqrt{x}$, respectively.
        The tube length is  $L=2000$ and the random walk probabilities as
        in \fig\ref{fig:profile_LD}.}
\label{fig:deltarhob_asep_r357}}
\end{center}
\end{figure}

These conclusions agree with the results of Monte Carlo simulations.
In these simulations, the tube length $L$ is necessarily finite.  This
implies that the profiles observed in the simulations may not reach
the asymptotic behavior present in a tube of infinite length. Since
both the crossover length $y_*$ and the extrapolation length $y_{**}$
increase quickly with increasing tube radius $R$, we expect to find
the true asymptotic behavior only for sufficiently small values of
$R$.  This expectation is confirmed by the simulation results.  For
sufficiently small $R$, the density deviation $\eta_\bd$ is found to
decay as $\sim 1/x^{1/2}$ for large $x$, as shown in
\fig\ref{fig:deltarhob_asep_r357} for $R=3$.  For sufficiently large
values of $R$, on the other hand, the observed profiles decay as
$1/x$, see the Monte Carlo data in \fig\ref{fig:deltarhob_asep_r357}
for $R=5$ and $R=7$.  In the latter cases, the true asymptotic
behavior is not accessible and is cut off by the finite value of $L$.

\begin{figure}[htbp]
\begin{center}
  \leavevmode
  \psfig{file=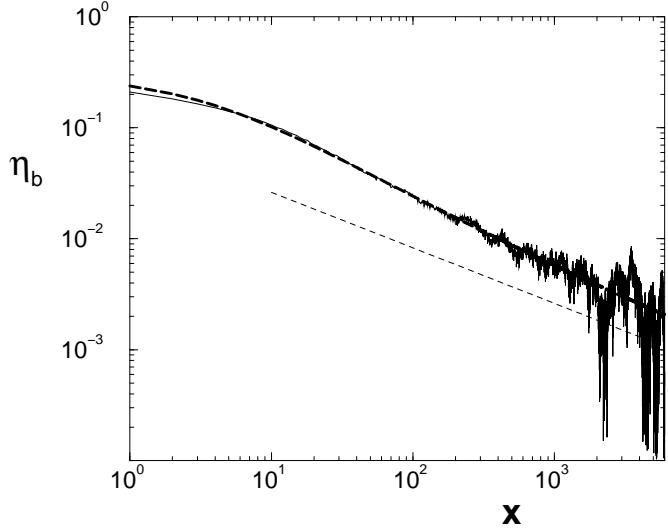,angle=-90,width=.5\textwidth}
  \parbox{\textwidth}{\caption{Deviation $\eta_\bd$ of the bound density
      from its constant value far from the boundaries as a function
      of the spatial coordinate $x$ for a tube with radius
      $R=3$ and length $L=6000$;  boundary conditions (B)  with
      $\rho_{\rm b,in}=0.8$ and $\rho_{\rm b,ex}=0.5$.  The thin
      solid line is obtained from the simulation, the thick dashed
      line is a fit to Eq.\ (\ref{E.bdensitydev}) with $D_{sc}\simeq 0.81$
      and $x_{sc}\simeq 8.57$. The
      dotted line indicates the power law $\sim  1/ \sqrt{ x}$. Note that the
      simulation data decay faster than $\sim 1/ \sqrt{ x}$ for
       $10^1 \protect\siml x \protect\siml 10^2$.  The random walk
      probabilities are the same as in \fig\ref{fig:profile_LD}.}
\label{fig:decay_L6000}}
\end{center}
\end{figure}

In addition, a more detailed analysis of the Monte Carlo results for
small values of $R$ but large values of $L$ explicitly shows that the
decay of the bound density deviation behaves as $\sim 1/\sqrt{x}$ for
large $x$ but decays faster than $1/\sqrt{x}$ for smaller values of
$x$. This crossover behavior is shown in \fig\ref{fig:decay_L6000}
where the simulation data are compared with the thin dashed line
corresponding to the decay law $\sim 1/\sqrt{x}$.  Furthermore, the
data can be well fitted with a density profile as given by
(\ref{E.bdensitydev}) if one makes an appropriate choice for the scale
parameters $D_{sc}$ and $x_{sc}$.

The Monte Carlo data shown in \fig\ref{fig:decay_L6000} correspond to
a tube of radius $R = 3$ and length $L = 6000$ with boundary densities
$\rho_{\rm b,in}=0.8$ and $\rho_{\rm b,ex}=0.5$.  In this case, a
least--squares fit of the data for $x > 20$ leads to the parameter
values $D_{sc}\simeq 0.81$ and $x_{sc}\simeq 8.57$.  The corresponding
fitting curve corresponds to the thick dashed line in
\fig\ref{fig:decay_L6000}.  The fit becomes less reliable close to the
left boundary, where the assumption that $\eta_\bd$ is small is no
longer fulfilled.  For fixed boundary densities, both fitting
parameters $D_{sc}$ and $x_{sc}$ are found to depend on $R$ and to
increase with increasing $R$.  In addition, for $R \ge 4$, it becomes
rather difficult to determine these parameters since one would have to
simulate rather long systems with $L \gg 6000$ in order to observe the
true asymptotic behavior.

For the ASEP in one dimension, it has been argued that the power--law
decay of the density profile as given by the asymptotic form $\approx
c/\sqrt{x/\ell}$ for large $x$ is characterized by the universal
amplitude $c = 1/2\sqrt{\pi} \simeq 0.282$. \cite{jans96,hage01}
Inspection of the relations (\ref{E.yDefined}), (\ref{E.aDefined}),
and (\ref{E.bdensitydev1}) shows that, in the present situation, $c =
D_{sc}/2 v_b \sqrt{x_{sc}}$ where the parameters $D_{sc}$ and $x_{sc}$
depend on the tube radius $R$. This implies that the amplitude $c$
will also depend on $R$. Using the numerically determined values for
$D_{sc}$ and $x_{sc}$ and the bound state velocity $v_b \simeq 0.99$,
we obtain the estimates $c \simeq 0.27$ for $R = 0$, which corresponds
to the 1--dimensional ASEP and should be compared with the exact value
$c = 1/2\sqrt{\pi} \simeq 0.282$, and $c \simeq 0.14$ for $R = 3$.
Thus, for small values of $R$, the amplitude $c$ is found to decrease
with increasing $R$.

\section{Diffusive injection and extraction of motors}
\label{sec:boundDiff}

Finally, we consider the second type of open boundary conditions
corresponding to case (C) in \fig\ref{fig:motor_tube_randbed}.  The
length of the tube is again denoted by $L$. This tube is now longer
than the filament which has length $L_{\rm F} < L$. The left end of
the tube is located at $x = 0$ as before but the left end of the
filament is at $x= \Delta L \equiv (L-L_{\rm F})/2$. Likewise, the
right end of the filament is at $x = \Delta L + L_{\rm F}$ whereas the
right end of the tube is at $x = 2 \Delta L + L_{\rm F} = L$.  Thus, a
motor particle which enters the tube on the left must diffuse over a
distance $ \sim \Delta L$ before it can come into contact with the
filament, and a motor particle which leaves the filament at its right
end must also diffuse over a distance $ \sim \Delta L$ before it can
leave the tube.

At the left and right end of the tube, we now prescribe constant
boundary densities as given by \be \rho(x=0, y,z) = \rho_{\rm ub,in}
\qquad {\rm and} \qquad \rho(x=L+1, y,z) = \rho_{\rm ub,ex} \quad \ee
for all values of $y$ and $z$ with $y^2 + z^2 \le R^2$.

As before, the jump probability $\beta$ to make backward steps on the
filament is taken to be zero. Likewise, the resting probability
$\gamma$ to make no step at all on the filament is also zero with a
possible exception at the 'last' filament site with $(x,y,z) = (
\Delta L + L_{\rm F}, 0, 0)$.  Indeed, in order to define the system
in a unique way, we still have to specify the probability to make a
forward step at this 'last' filament site.  Two possible choices
appear rather natural: (i) {\em Active} unbinding from the 'last'
filament site, i.e., the motor particle attempts to step forward with
probability $\alpha$ and makes a step if the adjacent nonfilament site
is unoccupied. In this case, the forward step at the last filament
site is governed by the same probabilities as all other forward steps
along the filament; and (ii) {\em Thermal} unbinding in which the
motor particle unbinds with probability $\epsilon/6$ both in the
forward direction and in the four orthogonal directions. In the latter
case, one has to choose the resting probability $\gamma$ to be nonzero
and to be given by $\gamma = 1 - 5 \epsilon/6$.

The choice (i) is suggested by the results of recent experiments on
microtubules and kinesin motors \cite{Surrey__Karsenti2001} which
indicate that these motors unbind quickly at the filament ends.  In
the following subsections, we will first consider this choice (i)
corresponding to active unbinding from the last filament site. In the
last subsection \ref{S.ActiveVersusThermal}, we will show that the
choice (ii) leads to rather similar behavior.

\subsection{Diffusive bottlenecks}
\label{S.Bottlenecks}

In order to understand the behavior found for boundary condition (C),
it is instructive to partition the tube into three compartments which
are defined as follows: (i) A left compartment with $1 \le x < \Delta
L$ where transport is purely diffusive; (ii) A middle compartment with
$\Delta L \le x \le \Delta L + L_{\rm F}$ where all directed (or
active) transport occurs; and (iii) A right compartment with $\Delta L
+ L_{\rm F} < x \le L$ where the transport is again purely diffusive.

For a stationary state, the total current through the tube must be
constant and, thus, must be the same in all three compartments. The
current through the middle compartment is given by the bound current
$j_\bd = v_\bd\rho_\bd^0(1-\rho_\bd^0)$.  Thus, the diffusive currents
$J_{\rm dif,L}$ and $J_{\rm dif,R}$ in the left and right tube segment
must be equal and must satisfy the simple relation
\begin{equation}
J_{\rm dif,L}=J_{\rm dif,R} =  v_\bd\rho_\bd^0(1-\rho_\bd^0)
\quad .
\label{E.CurrentCaseC}
\end{equation}

The relation as given by (\ref{E.CurrentCaseC}) is easily checked in
the simulations since the density profile is found to be approximately
linear in the left and in the right compartments provided $\Delta L$
is sufficiently large.  For such a linear density profile in the right
compartment, the diffusive current $J_{\rm dif,R}$ can be estimated as
\be J_{\rm dif,R} \simeq (1+N_{\rm ch}) D_\ub \frac{|\rho_\ub(x=\Delta
  L + L_{\rm F}) - \rho_{\rm ub, ex}|}{\Delta L} \quad .
\label{E.DiffusiveCurrent}
\ee Since the maximal density difference is one (in the units used
here), the maximal diffusive current behaves as \be \max(J_{\rm
  dif,R}) \sim (1+N_{\rm ch}) D_\ub \frac{1}{\Delta L} \sim R^2 D_\ub
\frac{1}{\Delta L} \quad .  \ee This shows that the maximal diffusive
current depends on the tube radius $R$, the lateral size $\Delta L$ of
the diffusive segments, and the diffusion coefficient $D_\ub $ of the
unbound motors. Since these parameters can be chosen independently
from the bound motor velocity $v_\bd$, the diffusive current $J_{\rm
  dif,L}$ can be made smaller than the maximal bound current $v_\bd/4$
on the filament. In the latter case, the diffusive compartments act as
diffusive bottlenecks and the maximal current phase characterized by
the current $v_\bd/4$ is expected to be absent from the phase diagram.
This expectation is indeed confirmed by the simulations as discussed
next.

\subsection{Phase diagram without maximal current phase}
\label{S.PDnoMCPhase}

\begin{figure}[htbp]
  \begin{center}
    \leavevmode
    \psfig{file=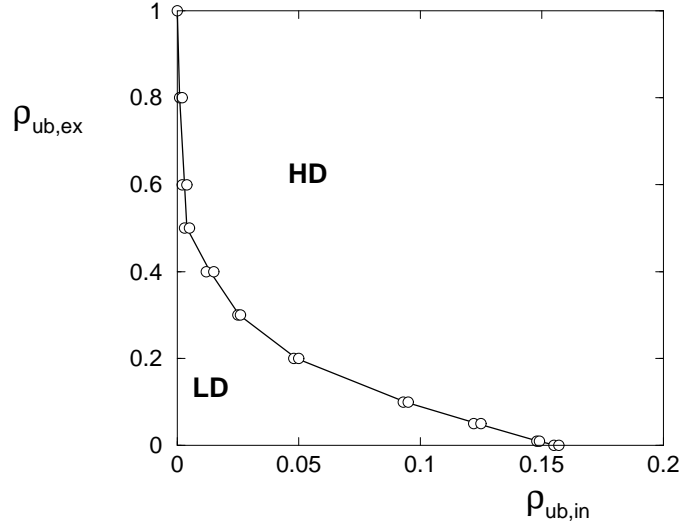,angle=-90,width=.5\textwidth}
    \caption{Phase diagram for tubes with boundary conditions (C) as
      a function of the left and right boundary densities
      $\rho_{\rm ub,in}$ and $\rho_{\rm ub,ex}$ for a
      set of parameters where no maximal current phase occurs. The tube has
      length $L=600$ and radius $R=5$, the filament length is $L_F=590$ and
      the distance between the filament ends and the tube ends is
      $\Delta L=5$. The random walk probabilities are the same
      as in \fig\ref{fig:profile_LD}. Note the different scales of the 
      axes.}
    \label{fig:phasendiagramm_R5b5_ausschnitt}
  \end{center}
\end{figure}

One geometry, for which no maximal current phase has been observed in
the simulations, is provided by a tube with radius $R = 5$, length $L
= 600$, and filament length $L_{\rm F} =590$.  The corresponding phase
diagram as determined by the Monte Carlo simulations is shown in
\fig\ref{fig:phasendiagramm_R5b5_ausschnitt}.  The largest part of the
phase diagram is covered by the high density phase; in addition, a low
density phase is found for small values of the boundary densities
$\rho_{\rm ub,in}$ and $\rho_{\rm ub,ex}$.

\begin{figure}[htbp]
  \begin{center}
    \leavevmode
    \psfig{file=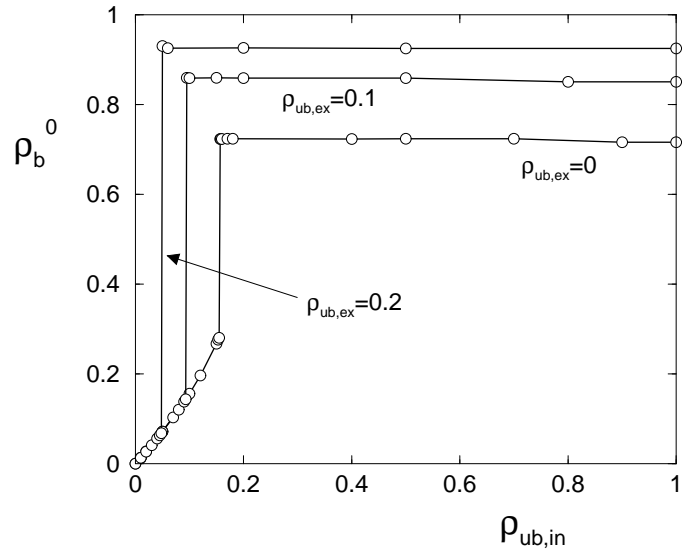,angle=-90,width=.5\textwidth}
    \caption{Bound density $\rho_\bd^0$ as a function of the left
      boundary density
      $\rho_{\rm ub,in}$ for boundary conditions (C) and
       different values of the right boundary
      density $\rho_{\rm ub,ex}$.
     The parameters are the same as in
     \fig\ref{fig:phasendiagramm_R5b5_ausschnitt}.
     The discontinuity in the functional dependence of $\rho_\bd^0$ on
     $\rho_{\rm ub,in}$ corresponds to the transition from the low to the
     high density phase, compare \fig\ref{fig:phasendiagramm_R5b5_ausschnitt}.}
    \label{fig:rho_b_ohneMC}
  \end{center}
\end{figure}

The transition line displayed in
\fig\ref{fig:phasendiagramm_R5b5_ausschnitt} has been determined from
the functional dependence of the bound density in the bulk,
$\rho_\bd^0$, on the left boundary density $\rho_{\rm ub,in}$ as shown
in \fig\ref{fig:rho_b_ohneMC}.  Inspection of this latter figure shows
that the bound bulk density $\rho_\bd^0$ jumps at a certain value of
the left boundary density $\rho_{\rm ub,in}$. Since $L$ is finite,
this jump occurs over a small but finite interval of $\rho_{\rm
  ub,in}$. Thus, the jump can be characterized by two $\rho_{\rm
  ub,in}$--values, say $\rho_{\rm ub,in}^<$ and $\rho_{\rm ub,in}^>$,
which represent the left and the right 'corner' of the numerically
determined jump. Both $\rho_{\rm ub,in}$--values have been included in
the phase diagram of \fig\ref{fig:phasendiagramm_R5b5_ausschnitt}.

For $\rho_{\rm ub,in}^< < \rho_{\rm ub,in} < \rho_{\rm ub,in}^>$, the
simulations do not reach a stationary state within two days of
computation. Simulations also become very slow in the high density
phase especially when the overall motor concentration gets so large
that the bound density is close to one and the unbound density is no
longer small compared to the bound density.

\subsection{Presence of the maximal current phase}
\label{S.CriteriaMCPhase}

In order to estimate the set of parameters, for which the phase
diagram exhibits a maximal current phase, we return to the estimate
(\ref{E.DiffusiveCurrent}) for the diffusive current $J_{\rm dif,R}$
and make the simplifying assumption that the bound and unbound
densities in the middle compartment of the tube are essentially
constant. Thus, we replace the true unbound density $\rho_\ub(x=\Delta
L + L_{\rm F})$ at the right end of the filament by its bulk value
$\rho_\ub^0$.

The bulk density $\rho_\ub^0$ of the unbound motors is related to the
bulk density $\rho_\bd^0$ of the bound motors via the radial
equilibrium relation (\ref{RDB_Formel}). For the maximal current
phase, one has $\rho_\bd^0 = 1/2$ and (\ref{RDB_Formel}) leads to
$\rho_\ub^0=\frac{\epsilon/\piad}{1+\epsilon/\piad} \approx
\epsilon/\piad$ for small $\epsilon$. In this way, we arrive at the
estimate \be J_{\rm dif,R} \simeq (1+N_{\rm ch}) D_\ub
\frac{\epsilon}{\piad \Delta L} \qquad {\rm for } \qquad \rho_{\rm ub,
  ex} = 0 \quad .
\label{E.Estimate2}
\ee

The maximal current phase should be present in the phase diagram if
the diffusive current $J_{\rm dif,R}$ exceeds the maximal current
$v_\bd/4$ on the filament. It then follows from (\ref{E.Estimate2})
that the maximal current phase should be present for $\rho_{\rm ub,
  ex} = 0$ provided the tube radius $R$ satisfies \be R^2 > R_*^2
\equiv \frac{v_\bd}{4 \pi} \frac{\piad \Delta L}{D_\ub \epsilon}
\label{E.PresenceMCPhase}
\ee where $1+N_{\rm ch} \approx \pi R^2$ has been used.  Using the
same line of reasoning, a second, less restrictive condition can be
obtained from an estimate for the diffusive current within the left
compartment.

The threshold value $R_*$ for the tube radius as given by
(\ref{E.PresenceMCPhase}) has been confirmed by Monte Carlo
simulations for the jump probabilities $\alpha = 1 - 2 \epsilon/3$
with $\epsilon = 1/100$, the sticking probability $\piad = 1$, the
compartment size $\Delta L = 5$, and the boundary densities $\rho_{\rm
  ub, in} = 0.2$ and $\rho_{\rm ub, ex} =0$.  The jump probabilities
imply the bound motor velocity $v_\bd = 1- 2/300$; the diffusion
coefficient $D_\ub$ of the unbound motors has the value $D_\ub = 1/6$
(since the resting probability $\gamma = 0$ as mentioned above). When
these parameter values are inserted into (\ref{E.PresenceMCPhase}),
one obtains the estimate $R_* \simeq 15.4$.

\begin{figure}[htbp]
  \begin{center}
    \leavevmode
    \psfig{file=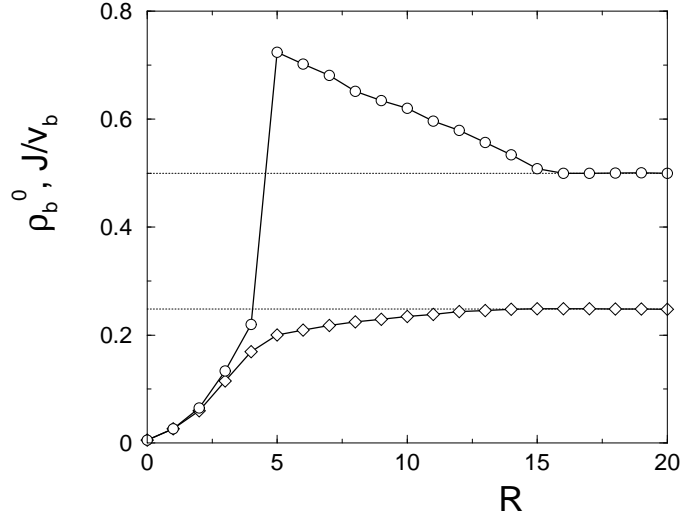,angle=-90,width=.5\textwidth}
    \caption{Current $J/v_\bd$ (diamonds) and bound density $\rho_\bd^0$
      (circles) as a function
      of the tube radius $R$ for boundary conditions (C) with
      $\rho_{\rm ub,in}=0.2$, $\rho_{\rm ub,ex}=0$.
      The tube length $L=600$,  the filament length   $L_F=590$,
      and the
      distance between the filament  ends and the tube  ends is
      $\Delta L=5$. The random walk probabilities are the same as
      in \fig\ref{fig:profile_LD}. }
    \label{fig:J_rhob_R_abh}
  \end{center}
\end{figure}

The corresponding Monte Carlo data are displayed in
\fig\ref{fig:J_rhob_R_abh}.  Inspection of this figure shows that the
current does indeed attain its maximal value $v_\bd/4$ for $R \geq
R_*$ with $R_* \simeq 16$.  The same \fig\ref{fig:J_rhob_R_abh} also
shows the transition from the low density to the high density phase
which occurs for a tube radius $R_{**}$ which satisfies $4 < R_{**} <
5$.

\begin{figure}[htbp]
  \begin{center}
    \leavevmode
    \psfig{file=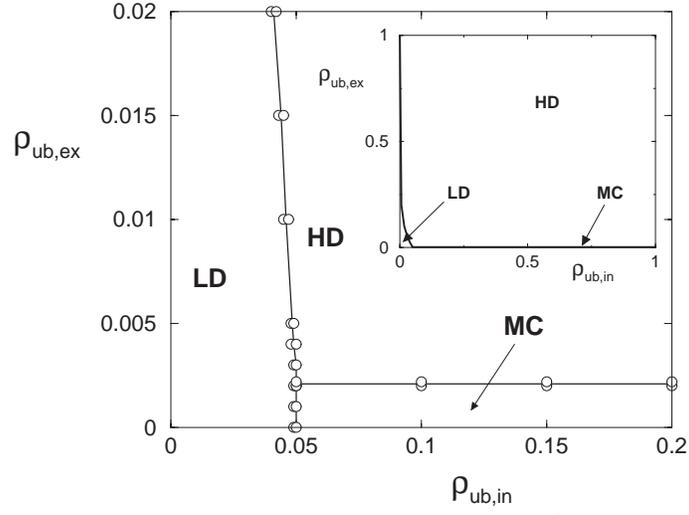,angle=-90,width=.5\textwidth}
    \caption{Phase diagram for a tube of radius $R=17$ with boundary
      conditions (C) as a function of the left and right boundary densities
      $\rho_{\rm ub,in}$ and $\rho_{\rm ub,ex}$.
      The other parameters are the same as in
      \fig\ref{fig:phasendiagramm_R5b5_ausschnitt}.
      In the inset, which shows the
      complete phase diagram, the maximal current phase can be hardly
      distinguished from the line $\rho_{\rm ub,ex}=0$.}
    \label{fig:phasendiagramm_R17b5}
  \end{center}
\end{figure}

A complete phase diagram for a tube with radius $R =17$ is shown in
\fig\ref{fig:phasendiagramm_R17b5}.  Again, most of the phase diagram
is covered by the high density phase (HD), while a low density phase
(LD) is found only for very small values of $\rho_{\rm ub,in}$.  The
maximal current phase (MC) is present now but only for very small
values of $\rho_{\rm ub,ex}$.

It is interesting to note that similar effects also occur in the
purely one-dimensional system if one considers a driven system which
is bounded by two segments which exhibit only diffusive transport.  In
this 1--dimensional case, quantitative predictions can be made using a
mean field approximation as will be discussed elsewhere
\cite{Klumpp_Lipowsky??}.

\subsection{Active versus thermal unbinding from the 'last' filament site}
\label{S.ActiveVersusThermal}

\begin{figure}[htbp]
  \begin{center}
    \leavevmode
    \psfig{file=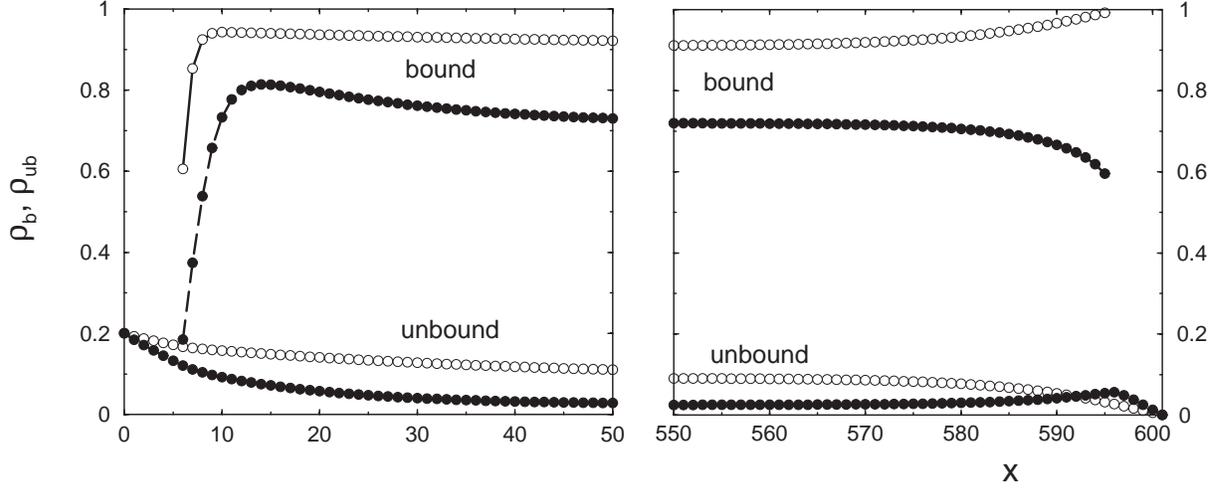,angle=-90,width=.9\textwidth}
    \caption{Bound and unbound density profiles $\rho_\bd$ and $\rho_\ub$
      as functions of the spatial coordinate $x$
      for boundary conditions (C) with $\rho_{\rm ub,in}=0.2$ and
      $\rho_{\rm ub,ex}=0$.
      Two different unbinding processes of the motor particles at the
      'last' filament site are compared: (Filled circles) Active unbinding
      for
      which the motor makes one final  step in the forward direction; and
      (Open Circles) Thermal unbinding with
      probability $\epsilon/6$ in the forward direction. The geometric
      parameters are $L=600$, $L_F=590$, $\Delta L=5$,  and $R=5$;   the
      random walk probabilities are as in \fig\ref{fig:profile_LD}.  }
    \label{fig:profile_abloesen_eps_v}
  \end{center}
\end{figure}

The Monte Carlo data displayed so far have been obtained for active
unbinding of a motor particle which is bound to the 'last' filament
site.  As mentioned above, another possibility is that the motor
particle gets stuck at the 'last' filament site and unbinds only by
thermal excitations, i.e., with unbinding probability $\epsilon/6$.
For these two different unbinding mechanisms, one will, in general,
obtain different density profiles.  However, this difference is not
dramatic as one can see from \fig\ref{fig:profile_abloesen_eps_v}
which exhibits density profiles for both cases. Although the
probability for a forward step at the 'last' filament site differs by
two orders of magnitude for the two cases, the bulk density exhibits a
relatively small difference. The bound density, on the other hand
increases and decreases close to the right end of the filament for
thermal and active unbinding from the 'last' site, respectively.

\section{Summary and conclusions}

Let us summarize the main results and add a few comments. We have
studied a lattice model for the motion of many molecular motors in an
open tube which contains a single filament. When bound to the
filament, the motor particles undergo an asymmetric simple exclusion
process (ASEP).  In addition, motors can unbind from the filament and
then diffuse freely in the tube.  As for the ASEP in one dimension, the
motor traffic in open tubes can exhibit three different phases: high
density and low density phases which are characterized by an
exponential decay of the density deviations from their bulk values
and maximal current phases characterized by an algebraic decay.
Therefore, the molecular motor traffic in open tubes are promising
candidates for the experimental observation of boundary--induced
non--equilibrium phase transitions.

In general, the location of the transition lines is found to depend on
the precise choice of the boundary conditions. Apart from periodic
boundary conditions, case (A), we studied two different boundary
conditions (B) and (C) for an open tube. In case (B), the bound and
unbound densities are kept fixed at the boundaries and satisfy radial
equilibrium. For this case, the location of the transition lines is
independent of the model parameters, and the phase diagram of the ASEP in
one dimension is recovered.  In case (C), the active compartment of
the tube is bounded by two compartments where the transport is purely
diffusive. In this latter case, the phase diagram depends on the
geometry of the tube and on the transport properties in the bound and
unbound motor states.  In many cases, the maximal current phase is
completely suppressed by the coupling to the diffusive compartments
which act as bottle necks for the transport.

The theoretical results described here should be accessible to
experiments on cytoskeletal filaments and motors.  In particular, the
motor traffic through open tubes as discussed here provides new
opportunities to study the transport properties of ASEPs by systematic
experiments.

\section*{Acknowledgments}

We thank Theo M. Nieuwenhuizen for comments and discussions during the
initial stage of this work.


\appendix

\section{Radial equilibrium for periodic  boundary conditions}
\label{app_RDB}

In the following, we show that \eq(\ref{RDB_Formel}), the condition
for radial equilibrium, holds exactly in the case of periodic boundary
conditions using the quantum Hamiltonian formalism \cite{Schuetz2001}.
The exact stationary master equation can be written in the form
\begin{equation}
  H|\rho\rangle=0,
\end{equation}
where $H$ is the 'quantum Hamiltonian' of the stochastic process and
$|\rho\rangle$ is a vector in a product Hilbert space; each lattice
site is represented by a two-state system with the orthogonal vectors
$|1\rangle$ for an occupied lattice site ("spin down") and $|0\rangle$
for a vacancy ("spin up").  Because of translational invariance in the
direction parallel to the filament, there cannot be any radial
currents in our system, and the unbound density is independent of the
radial coordinate; thus we can restrict the analysis to the case of
one unbound channel.  We denote by $|\rho\rangle_{k,\bd}$ and
$|\rho\rangle_{k,\ub}$ the state of site $k$ of the bound and unbound
channel, respectively.  Using the general recipe given in chapter 2 of
Ref.\ \cite{Schuetz2001} we construct the 'quantum Hamiltonian' $H$ of
our system:
\begin{equation}
  H = H_1 + H_2 +H_3,
\end{equation}
where $H_1$ represents the dynamics of the asymmetric exclusion
process in the bound channel, $H_2$ the symmetric exclusion process in
the unbound channel and $H_3$ the coupling of the two channels. Each
term can be written as a sum $H_i=\sum_k h_k^{(i)}$ with
\begin{eqnarray}
  h_k^{(1)} & = & v_\bd \left(
n_{k,\bd}v_{k+1,\bd}-s_{k,\bd}^+s_{k+1,\bd}^- \right) \\
  h_k^{(2)} & = & D_\ub \left(
n_{k,\ub}v_{k+1,\ub}-s_{k,\ub}^+s_{k+1,\ub}^-
+v_{k,\ub}n_{k+1,\ub}-s_{k,\ub}^-s_{k+1,\ub}^+  \right) \\
  h_k^{(3)} & = & \frac{\epsilon}{6} \left( n_{k,\bd}v_{k,\ub}
-s_{k,\bd}^+s_{k,\ub}^-\right)+\frac{\piad}{6} \left( v_{k,\bd}n_{k,\ub}
-s_{k,\bd}^-s_{k,\ub}^+\right).
\end{eqnarray}
$s_{k,\bd}^-$ is a creation operator for a particle at site $k$ of the
bound channel, and $s_{k,\bd}^+$ is the corresponding annihilation
operator. $n_{k,\bd}$ is the particle number operator at site $k$ of
the bound channel and $v_{k,\bd}=1-n_{k,\bd}$.  Operators for the
unbound channel are defined in an analogous manner.  The off-diagonal
parts of the operators $h_k^{(i)}$ represent hopping of particles,
while the diagonal parts are determined by conservation of
probability, see chapter 2 of Ref.\ \cite{Schuetz2001}.

We now show that the product measure
\begin{equation}\label{productMeasure}
  |\rho\rangle=\bigotimes_k \Big( \left[ (1-\rho_\bd)|0\rangle_{k,\bd}
+\rho_\bd|1\rangle_{k,\bd}\right] \otimes \left[
(1-\rho_\ub)|0\rangle_{k,\ub} +\rho_\ub|1\rangle_{k,\ub}\right]\Big),
\end{equation}
is a stationary state, if \eq(\ref{RDB_Formel}) holds, i.e.\ that the
radial equilibrium condition implies $H|\rho\rangle=0$.  The product
measure $|\rho\rangle$ defines a state where the density is $\rho_\bd$
at each lattice site of the bound channel and $\rho_\ub$ at each site
of the unbound channel and spatial correlations vanish.

$H_1$ and $H_2$ do not couple the bound and unbound channel, therefore
we can consider them separately and refer to the result, that for the
symmetric as well as for the asymmetric exclusion process, the product
measure $|\rho\rangle$ is stationary in the case of periodic boundary
conditions \cite{Spitzer1970}.  For a proof using the quantum
Hamiltonian formalism, see chapter 7.1.2 of Ref.\ \cite{Schuetz2001}:
In the case of the ASEP in one dimension, for example, one can easily
check, that
$h_{k}^{(1)}|\rho\rangle=v_\bd(n_{k,\bd}-n_{k+1,\bd})|\rho\rangle$,
therefore the summation gives zero for periodic boundary conditions.
Hence $H_1|\rho\rangle=0$ and $H_2|\rho\rangle=0$.

Concerning $H_3$, it is sufficient to consider a single site $k$ in
both channels.  Doing some steps, we find
\begin{equation}
  \left( n_{k,\bd}v_{k,\ub}
-s_{k,\bd}^+s_{k,\ub}^-\right)|\rho\rangle_{k,\bd}|\rho\rangle_{k,\ub}=\rho_\bd(
1-\rho_\ub)\Big(|1\rangle_{k,\bd}|0\rangle_{k,\ub}-|0\rangle_{k,\bd}|1\rangle_{k
,\ub}\Big)
\end{equation}
and
\begin{equation}
 \left( v_{k,\bd}n_{k,\ub} -s_{k,\bd}^-s_{k,\ub}^+\right)
 |\rho\rangle_{k,\bd}|\rho\rangle_{k,\ub}=\rho_\ub(1-\rho_\bd)
 \Big(|0\rangle_{k,\bd}|1\rangle_{k,\ub}-|1\rangle_{k,\bd}|0\rangle_{k,\ub}
 \Big)
\end{equation}
for the product measure (\ref{productMeasure}). Hence from
\begin{equation}
  h_k^{(3)}|\rho\rangle_{k,\bd}|\rho\rangle_{k,\ub}=\left(\frac{\epsilon}{6}
  \rho_\bd(1-\rho_\ub)-\frac{\piad}{6}\rho_\ub(1-\rho_\bd)\right)
  \Big(|1\rangle_{k,\bd}|0\rangle_{k,\ub}-|0\rangle_{k,\bd}|1\rangle_{k,\ub}
  \Big)=0,
\end{equation}
we obtain \eq (\ref{RDB_Formel}), which is the condition for radial
equilibrium.

\section{Continuum two-state mean field equations for open boundaries}
\label{app_mf}

In this appendix, we give the details of the continuum mean field
approximation for the two--state model introduced in section
\ref{sec:profile}.  We insert the decomposition
$\rho_\bd=\rho_\bd^0+\eta_\bd $ and $\rho_\ub=\rho_\ub^0+\eta_\ub$ as
introduced in (\ref{E.DensityDecomposition}) into the mean field
equations (\ref{mf1}) and (\ref{mf2}) and expand these equations up to
second order in the density deviations $\eta_\bd$ and $\eta_\ub$. As a
result, we obtain
\begin{equation}
\label{mf1_entw}
  D_\bd \dx\eta_\bd + N_{\rm ch} D_\ub \dx\eta_\ub= v_\bd(1-2\rho_\bd^0)
\eta_\bd-v_\bd\eta_\bd^2
\end{equation}
and
\begin{equation}
\label{mf2_entw}
  v_\bd(1-2\rho_\bd^0)\dx\eta_\bd-2v_\bd\eta_\bd\dx\eta_\bd-D_\bd
\frac{\partial^2}{\partial x^2}\eta_\bd= A\eta_\ub -B\eta_\bd
+(\tilde\epsilon-\tpi)\eta_\ub\eta_\bd,
\end{equation}
with
\begin{equation}\label{def_AB}
A\equiv\tpi(1-\rho_\bd^0)+\tilde\epsilon\rho_\bd^0
\qquad {\rm and} \qquad B\equiv\tilde\epsilon(1-\rho_\ub^0)+\tpi\rho_\ub^0.
\end{equation}
Note that radial equilibrium for the densities $\rho_\bd$ and
$\rho_\ub$ would imply that the right hand side of (\ref{mf2_entw})
vanishes.

Let us first consider the case $\rho_\bd^0\neq 1/2$ which applies to
the high and low density phases. In this case, we can neglect the
second order terms and obtain two equations which are linear in the
density deviations.  The second equation (\ref{mf2_entw}) can then be
solved for $\eta_\ub$ which leads to
\begin{equation}
  \eta_\ub=g \,\eta_\bd+\frac{v_\bd(1-2\rho_\bd^0)}{A}\dx\eta_\bd-
\frac{D_\bd}{A}\frac{\partial^2}{\partial x^2}\eta_\bd
\end{equation}
with
\begin{equation}
g \equiv  \frac{B}{A} = \frac{\tilde\epsilon(1-\rho_\ub^0)+\tpi\rho_\ub^0}
{\tpi(1-\rho_\bd^0)+\tilde\epsilon\rho_\bd^0}
\quad .
\end{equation}
When this expression is inserted into (\ref{mf1_entw}), we obtain the
first order relation
\begin{equation}
\label{hdld_gl}
  (D_\bd+ N_{\rm ch} D_\ub g )\dx\eta_\bd+N_{\rm ch}
D_\ub\frac{v_\bd(1-2\rho_\bd^0)}{A} \frac{\partial^2}{\partial x^2}
\eta_\bd -N_{\rm
ch} D_\ub\frac{D_\bd}{A}\frac{\partial^3}{\partial
x^3}\eta_\bd=v_\bd(1-2\rho_\bd^0)\eta_\bd.
\end{equation}
for $\eta_\bd$.  We now make the Ansatz $\eta_\bd\sim {\rm
  exp}(x/\xi)$ which leads to the cubic equation (\ref{E.xi}) for the
decay length $\xi$.

Now we consider the maximal current phase, i.e.\ the case
$\rho_\bd^0=1/2$. In this case the linear terms are zero. Furthermore,
we neglect terms of order $\eta_\bd\frac{\partial \eta_\bd}{\partial
  x}$ and $\frac{\partial^2 \eta_\bd}{\partial x^2}$ as can be
justified a posteriori since $\eta_\bd$ is found to decay as an
inverse power of $x$.  Thus, up to leading order, we can ignore the
left hand side of (\ref{mf2_entw}), and the right hand side of
(\ref{mf2_entw}) vanishes. This implies radial equilibrium for the
asymptotic decay to the homogeneous solution. Up to this order, we
find
\begin{equation}
  \eta_\ub=\frac{B\eta_\bd}{A+(\tilde\epsilon-\tpi)\eta_\bd} \approx
g \eta_\bd-\frac{B(\tilde\epsilon-\tpi)}{A^2}\eta_\bd^2
\end{equation}
for small $\eta_\bd$ and therefore
\begin{equation}
  \dx\eta_\ub = g \dx\eta_\bd+O(\eta_\bd\frac{\partial
\eta_\bd}{\partial x}).
\end{equation}
If this latter expression is inserted into (\ref{mf1_entw}), we obtain
\begin{equation}
  \label{mf_rhobd}
  (D_\bd+ g N_{\rm ch} D_\ub)\dx\eta_\bd \approx -v_\bd\eta_\bd^2
\end{equation}
which leads to
\begin{equation}
  \eta_\bd \approx \frac{D_\bd+ g \, N_{\rm ch} D_\ub}{v_\bd\,x}
\qquad {\rm for } \quad {\rm large} \quad x \quad .
\end{equation}
For $N_{\rm ch}=0$, this asymptotic behavior is identical to the one
found from the mean field approximation for the ASEP in one
dimension\cite{Krug1991}.

\newpage

\newpage

\section{List of symbols}

\begin{tabbing}
  symbol\quad\= definition\= \kill

  $a$ \> abbreviation for $  D_{sc} / 2 v_b x_{sc}$ used in subsection
  \ref{sec:profileMC} \\
  (A) \> periodic boundary conditions, see
  \fig\ref{fig:motor_tube_randbed}\\
  $A$\> abbreviation used in Appendix \ref{app_mf}, defined in
  \eq(\ref{def_AB})\\
  $\alpha$ \> probability for a forward step of the bound motor\\
  $b$ \> abbreviation for $1 + g \,N_{\rm ch} D_\ub/D_{sc}$
  used in subsection  \ref{sec:profileMC}\\
  (B) \> open tube, radial equilibrium at the boundaries,  see
  \fig\ref{fig:motor_tube_randbed}\\
  $B$\> abbreviation used in Appendix \ref{app_mf}, defined in
  \eq(\ref{def_AB})\\
  $\beta$ \> probability for a backward step of the bound motor\\
  (C) \> open tube, diffusive injection and extraction of motors, see
  \fig\ref{fig:motor_tube_randbed} \\

  $D$ \> diffusion coefficient for the 1--dimensional ASEP\\
  $D_\bd$ \> diffusion coefficient of the bound motor particle\\
  $\Delta L$ \> distance between filament end and tube end for boundary
conditions
  (C)\\
  $D_\ub$ \> diffusion coefficient of the unbound motor particle\\
  $D_{sc}$ \> scale parameter for the scale-dependent diffusion coefficient\\

  $\epsilon$    \>  unbinding probability of a bound motor \\
  $\tilde\epsilon$ \> rescaled unbinding probability
  $2\epsilon/3$ \\
  $\eta_\bd(x)$ \> deviation of the bound density from the constant bulk
  value\\
  $\eta_\ub(x)$ \> deviation of the unbound density from the constant bulk
value\\

  $g$ \> parameter defined by \eq(\ref{def_Gamma}) \\
  $\gamma$ \> resting probability of a bound motor\\

  $J$ \> (global) current through the tube\\
  $j_\bd$ \> local bound current\\
  $j_\ub$ \> local unbound current\\
  $J_{\rm dif,L}$ \> diffusive current in the left diffusive compartment for
  boundary conditions (C) \\
  $J_{\rm dif,R}$ \> diffusive current in the right diffusive compartment for
  boundary conditions (C)\\

  $L$ \> length of the tube\\
  $\ell$ \> lattice constant on filament\\
  $L_F$ \> length of the filament\\

  $N_{\rm ch}$ \> number of unbound channels\\
  $N_{\rm mo}$ \> number of motors in the tube which is fixed for periodic
boundary
  conditions (A)\\
  $\phi$ \> cross section of the tube\\
  $\piad$ \> sticking probability for a motor hopping to the filament\\
  $\tpi$ \> rescaled sticking probability $ 2\piad/3$\\
  $r$ \> radial spatial coordinate \\
  $R$ \> radius of the tube\\
  $R_*$ \> minimal tube radius for  which a maximal current phase is found
in case (C)\\

  $\rho(x)$ \> density in the case of the 1--dimensional ASEP\\
  $\rho_\bd(x)$ \>  density of motors bound to the filament\\
  $\rho_{\rm b,ex}$ \> right boundary density on the filament\\
  $\rho_{\rm b,in}$ \> left boundary density on the filament\\
  $\rho_\bd^0$ \> constant bulk density on the filament\\
  $\rho_{\rm ex}$ \> right boundary density for the 1--dimensional ASEP\\
  $\rho_{\rm in}$ \> left boundary density for the 1--dimensional ASEP\\
  $\rho_{\rm mo}$ \> overall   motor concentration\\
  $\rho^0$ \> constant bulk density for the 1--dimensional ASEP\\
  $\rho_\ub(x,r)$ \>  density of unbound motors\\
  $\rho_{\rm ub,ex}$ \>   density of unbound motors at the right boundary\\
  $\rho_{\rm ub,in}$ \>  density of unbound motors at the left boundary\\
  $\rho_\ub^0$ \> unbound constant bulk density\\

  $\tau$ \> basic time unit\\

  $v$ \> velocity for the 1--dimensional ASEP\\
  $v_\bd$ \> velocity of bound motor\\
  $v_c$ \> velocity of density fluctuations\\
  $v_s$ \> domain wall velocity\\

  $x$ \> spatial coordinate parallel to the filament\\
  $x_{sc}$ \> scale parameter for the scale-dependent
  diffusion coefficient, see \eq (\ref{scale_dep_D_b})\\
  $x_*,x_{**}$ \> crossover lengths\\
  $\xi$ \> decay length for the density deviations\\
  $\xi_0$ \> localization length for the 1-dimensional ASEP\\
  $x_0$ \> spatial location of the left tube end\\

  $y$ \>  rescaled and shifted spatial coordinate,  $y \equiv (x - x_0) /
x_{sc}$\\
  $y_*,y_{**}$ \> rescaled crossover lengths
\end{tabbing}


\newpage

\begin{figure}[htbp]
  \begin{center}
    \leavevmode
    \psfig{file=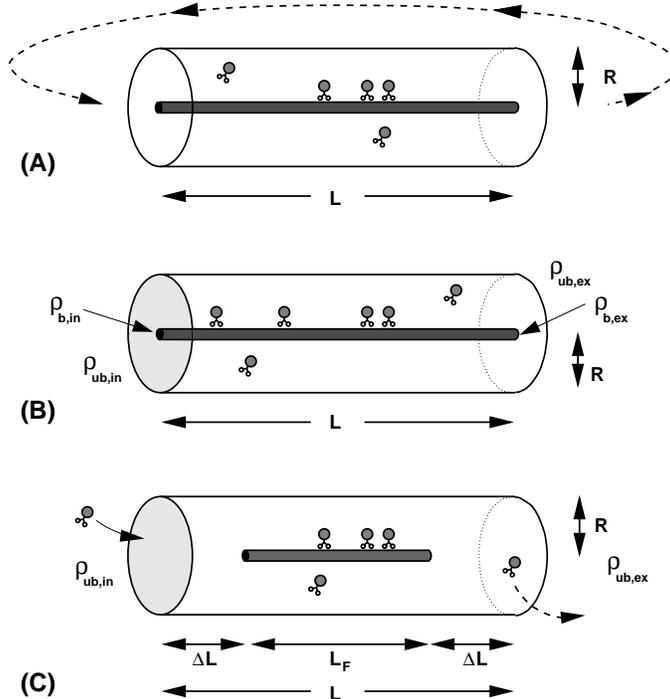,angle=0,width=.5\textwidth}
    \caption{Different types of  boundary conditions: (A) Periodic
      boundary conditions which is similar to a closed torus geometry; (B)
      Open tube with boundaries satisfying
      radial equilibrium:  the bound and unbound  motor densities are fixed at
      the two boundaries and satisfy radial detailed balance at each
      boundary; and (C) Open
      tube with diffusive injection and extraction of motors. In all cases,
      the tube has total length $L$; in cases (A) and (B), the filament  has
      the same length as the tube; in case (C), the filament
      has length $L_{\rm F} < L$ and there are two boundary compartments of
      linear size $\Delta L$. }
    \label{fig:motor_tube_randbed}
  \end{center}
\end{figure}

\begin{figure}[tbp]
\begin{center}
  \leavevmode
  \psfig{file=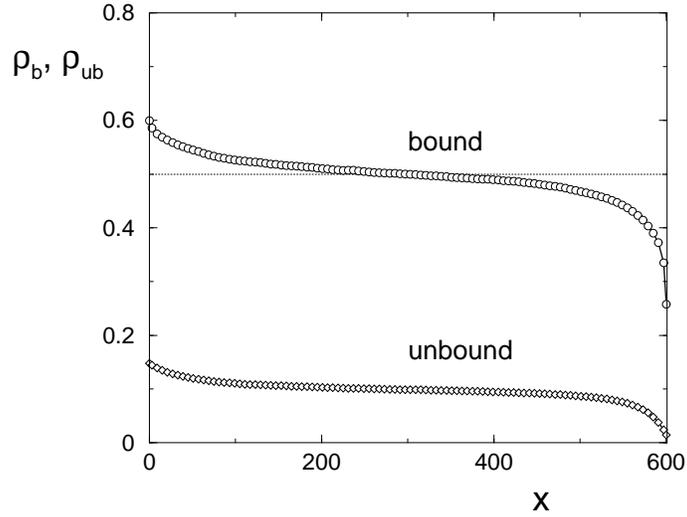,angle=-90,width=.5\textwidth}
  \parbox{\textwidth}{\caption{Bound and unbound density profiles
      $\rho_\bd$ and $\rho_\ub$
      as functions of the spatial coordinate $x$ for the maximal current
      phase; boundary conditions (B) with $\rho_{\rm b,in}=0.6$ and
      $\rho_{\rm b,ex}=0.1$.
      The unbound density has been averaged over the
      tube cross section and multiplied by a scale factor of 10. The
      parameters
      are the same as in \fig\ref{fig:profile_LD}.}
\label{fig:profile_MC}}
\end{center}
\end{figure}

\newpage

\end{document}